\documentclass[a4paper,headings=standardclasses]{scrartcl}
\pdfoutput=1
\usepackage[utf8]{inputenc}

\usepackage[T1]{fontenc} 
\usepackage{textcomp} 
\usepackage{lmodern} 
\usepackage[osf]{mathpazo} 
\usepackage{amsmath,amssymb}
\usepackage{mathtools} 
\usepackage{tensor}

\usepackage{tikz}
\usetikzlibrary{arrows.meta,decorations.markings}

\usepackage{todonotes}

\usepackage{enumitem}
\usepackage{relsize}
\newcommand{\acronym}[1]{\texorpdfstring{\textsmaller{#1}}{#1}}

\usepackage[british]{babel}

\usepackage{microtype} 

\usepackage{authblk}

\usepackage[colorlinks]{hyperref}
\usepackage[noabbrev]{cleveref}

\usepackage[numbers,sort&compress]{natbib}
\newcommand{\vect}[1]{\boldsymbol{#1}} 
\DeclareMathOperator{\Or}{O} 
\newcommand{\e}{\mathrm{e}} 
\newcommand{\I}{\mathrm{i}} 
\newcommand{\D}{\mathrm{d}} 
\newcommand{\R}{\mathbb R} 
\newcommand{\C}{\mathbb C} 

\numberwithin{equation}{section}

\title{Geometric post-Newtonian description of massive spin-half
  particles in curved spacetime}

\author[1,2,a]{Ashkan Alibabaei}
\author[1,b]{Philip K. Schwartz}
\author[1,3,c]{Domenico Giulini}
\affil[1]{Institute for Theoretical Physics,
  Leibniz University Hannover, \par
  Appelstraße 2, 30167 Hannover, Germany}
\affil[2]{Institute of Quantum Optics,
  Leibniz University Hannover, \par
  Welfengarten 1, 30167 Hannover, Germany}
\affil[3]{Center of Applied Space Technology and Microgravity,
  University of Bremen, \par
  Am Fallturm 1, 28359 Bremen, Germany}
\affil[a]{\normalfont\texttt{\href{mailto:aa2377@cantab.ac.uk}
    {aa2377@cantab.ac.uk}}}
\affil[b]{\normalfont\texttt{\href{mailto:philip.schwartz@itp.uni-hannover.de}
    {philip.schwartz@itp.uni-hannover.de}}}
\affil[c]{\normalfont\texttt{\href{mailto:giulini@itp.uni-hannover.de}
    {giulini@itp.uni-hannover.de}}}

\date{}

\begin{document}
\maketitle

\begin{abstract}
  \noindent
  We consider the Dirac equation coupled to an external
  electromagnetic field in curved four-dimensional spacetime with a
  given timelike worldline $\gamma$ representing a classical clock.
  We use generalised Fermi normal coordinates in a tubular
  neighbourhood of $\gamma$ and expand the Dirac equation up to, and
  including, the second order in the dimensionless parameter given by
  the ratio of the geodesic distance to the radii defined by spacetime
  curvature, linear acceleration of $\gamma$, and angular velocity of
  rotation of the employed spatial reference frame along $\gamma$.
  With respect to the time measured by the clock $\gamma$, we compute
  the Dirac Hamiltonian to that order.  On top of this `weak-gravity'
  expansion we then perform a post-Newtonian expansion up to, and
  including, the second order of $1/c$, corresponding to a
  `slow-velocity' expansion with respect to $\gamma$.  As a result of
  these combined expansions we give the weak-gravity post-Newtonian
  expression for the Pauli Hamiltonian of a spin-half particle in an
  external electromagnetic field.  This extends and partially corrects
  recent results from the literature, which we discuss and compare in
  some detail.
\end{abstract}

\section{Introduction}
\label{sec:Intro}

Modern experiments allow to probe the interface between quantum and
gravitational physics at a rapidly growing degree of accuracy.  A
proper theoretical description of such experiments would ideally be
based on a higher-level theory encompassing Quantum Mechanics as well
as General Relativity as appropriate limiting cases.  However, as is
well known, such a higher-level theory is still elusive.  Hence, we
cannot simply `compute' the impact of a classical gravitational field,
described by a (generally curved) spacetime metric, upon the dynamics
of a quantum system.  Rather, depending on the context, we must
`deduce' the influence of the gravitational field on the dynamics of
the quantum system from general principles that we expect to be robust
and eventually realised in the higher-level theory.  This is, in a
nutshell, the generally accepted strategy today for exploring the
interface between quantum and gravitational physics, to which the
present study also subscribes.

On the one hand, exploration of this interface is, of course, of
fundamental theoretical interest, not least since by such exploration
one hopes to gain insight into how to combine gravitational and
quantum physics on a broader level.  On the other hand, a systematic
understanding of this interface is also necessary in view of ongoing
progress in quantum-mechanical experiments, whose increasing accuracy
makes the consideration of post-Newtonian gravitational effects
inevitable.  As recent examples of interest in `novel' (i.e.\
previously not considered) gravitational effects in experiments, we
mention the question of the gravitational contribution to
high-precision measurements of the $g$-factor of an electron stored in
a Penning trap (also referred to as a `geonium atom') \cite
{Laszlo.Zimboras:2018,Jentschura:2018,Ulbricht.Mueller.Surzhykov:2019,
  Ito:2021}, and the recent results of $q$\textsc{Bounce}, which is a
Ramsey-type gravitational resonance spectroscopy experiment using
ultra-cold neutrons to test the neutron's coupling to the
gravitational field of the earth in the micrometer range
\cite{Micko.EtAl:2022}.

In the field of matter-wave interferometry, post-Newtonian
gravitational effects have recently even become a direct object of
investigation, expected to be observed in the foreseeable future with
the current generation of matter-wave interferometers
\cite{Asenbaum.EtAl:2017, Lezeik.EtAl:2022}: for systems possessing
internal degrees of freedom, post-Newtonian effects are expected to
induce a coupling between these internal degrees of freedom and the
system's external degrees of freedom \cite{Zych.EtAl:2011,
  Pikovski.EtAl:2015, Schwartz.Giulini:2019:AiG, Schwartz:2020}.  In
interferometry with such systems, dubbed `quantum clock
interferometry', these couplings may be observed and/or exploited for,
e.g., tests of (certain aspects of) the equivalence principle
\cite{Loriani.EtAl:2019,Roura:2020,Roura.EtAl:2021}.

In such examples, and in the more general context of gravitational
effects in quantum systems, it is important to base one's estimates of
possible gravity effects on a \emph{well-defined} and
\emph{systematic} approximation scheme.  Without such a controlled
scheme, a deviation of experimental observations from expectations
might be either (a) a result of the underlying theory being indeed
`wrong' (in the appropriate sense), or (b) simply an artefact of an
unsystematic way of deriving the alleged `theoretical predictions'.
That is: only by employing a consistent and systematic scheme one can
guarantee a complete and redundancy-free account of the (in our case
relativistic) corrections that one derives, as a necessary condition
for properly testing the underlying theory.

It is the aim of this paper to present such a scheme for a massive
spin-half particle obeying the Dirac equation in curved spacetime.
Our scheme is based on the assumed existence of a distinguished
reference wordline $\gamma$, which, e.g., may be thought of as that of
a clock in the laboratory or a distinguished particle.  In a tubular
neighbourhood of $\gamma$ we use generalised Fermi normal coordinates
\cite{Manasse.Misner:1963, Ni.Zimmermann:1978, Li.Ni:1979} with
reference to $\gamma$ and an adapted (meaning the unit timelike vector
is parallel to the tangent of $\gamma$) orthonormal frame along it.
The coordinates are `generalised' in the sense that we will allow the
worldline $\gamma$ to be accelerated, and the orthonormal frame to
rotate, i.e.\ its Fermi--Walker derivative need not vanish.  The
approximation procedure then consists of two steps which are logically
independent \emph{a~priori}.

In the first step we perform a `weak-gravity expansion', which means
that we expand the fields in the tubular neighbourhood of $\gamma$ in
terms of a dimensionless parameter given by the ratio of the spacelike
geodesic distance to $\gamma$ to the radii that are defined by
spacetime curvature, acceleration of $\gamma$, and the angular
velocity of rotation of the chosen frame along $\gamma$.  We recall
that the radius associated with $\gamma$'s acceleration $a$ is given
by $c^2/a$ and that the radius associated with the frame's angular
velocity $\omega$ (against a Fermi--Walker transported one) is
$c/\omega$.  The curvature radius is given by the inverse of the
modulus of the typical Riemann-tensor components with respect to the
orthonormal frame.  As first derivatives of the Riemann-tensor will
also appear, we also need to control these against \emph{third} powers
of the geodesic distance.  Our expansion hypotheses are summarised in
the expressions \eqref{eq:FNC_small_params}.  Consistently performing
this expansion is the content of \cref{sec:expansion_FNC}, leading to
the Dirac Hamiltonian \eqref{eq:Dirac_FNC}, which is our first main
result.  Note that a `Hamiltonian' refers to a `time' with respect to
which it generates the evolution of the dynamical quantities.  In our
case, that time is given by the proper time along $\gamma$, i.e.\ time
read by the `clock', extended to the tubular neighbourhood along
spacelike geodesics.

In the second step we perform a `slow-velocity' expansion by means of
a formal power series expansion in terms of $1/c$, i.e.\ a
\emph{post-Newtonian expansion}.  More specifically, we will expand
positive-frequency solutions of the (classical) Dirac equation as
formal power series in $c^{-1}$, similar to the corresponding
expansion for the Klein--Gordon equation as discussed in, e.g.,
\cite{Giulini.Grossardt:2012,Schwartz.Giulini:2019:PNSE,Schwartz:2020},
and in a broader context in \cite{Giulini.Grossardt.Schwartz:2022}.
For the case of $\gamma$ being a stationary worldline in a stationary
spacetime, this expansion may be considered a post-Newtonian
description of the one-particle sector of the massive Dirac quantum
field theory.  \emph{A priori} this `slow-velocity' approximation is
an independent expansion on top of the former.  But for the system
moving under the influence of the gravitational field the latter
approximation is only consistent with the former if the relative
acceleration of the system against the reference set by $\gamma$ stays
bounded as $1/c \to 0$.  This implies that the curvature tensor
components with respect to the adapted orthonormal frame should be
considered as being of order $c^{-2}$.  The coupled expansions then
lead us to the Pauli Hamiltonian \eqref{eq:Pauli_PN}, which is the
second main result of our paper.

Clearly, our work should be considered in the context of previous work
by others.  In 1980, Parker \cite{Parker:1980:PRL,Parker:1980:PRD}
presented explicit expressions for the energy shifts suffered by a
one-electron atom in free fall within a general gravitational field,
the only restriction imposed on the latter being that its time-rate of
change be sufficiently small so as to allow stationary atomic states
and hence well-defined energy levels.  Parker also used Fermi normal
coordinates, though standard ones, i.e.\ with respect to
\emph{non-rotating} frames along a \emph{geodesic} curve $\gamma$.  He
then gave an explicit expression for the Dirac Hamiltonian to what he
calls `first order in the [dimensionful] curvature', which in our
language means second order in the dimensionless ratio of geodesic
distance to curvature radius.  Regarding the `slow-velocity'
approximation, Parker considers only the leading-order terms, i.e.\
the Newtonian limit instead of a post-Newtonian expansion.

The restriction to non-rotating frames along $\gamma$ and geodesic
$\gamma$ was lifted by Ito \cite{Ito:2021} in 2021, who aimed for
estimating the inertial and gravitational effects upon $g$-factor
measurements of a Dirac particle in a Penning trap.  To that end he
presented an expansion in generalised Fermi normal coordinates of the
Dirac Hamiltonian also including terms to second order in the ratio
$\text{(geodesic distance)}/\text{(curvature radius)}$, but only to
first order in the ratios $\text{(geodesic
  distance)}/\text{(acceleration radius)}$, where `acceleration
radius' refers to both acceleration of $\gamma$ and the rotation of
the frames along $\gamma$ as explained above.  Ito also considers a
`non-relativistic limit' by performing a Fouldy--Wouthuysen
transformation \cite{Foldy.Wouthuysen:1950} with a transformation
operator expanded as a formal power series in $1/m$ (the inverse mass
of the fermionic particle).  In dimensionless terms, the latter
corresponds to a simultaneous expansion in $v/c$ as well as the ratio
$\text{(Compton wavelength)}/\text{(geodesic distance)}$.

Finally we mention the work of Perche \& Neuser
\cite{Perche.Neuser:2021} from 2021, who generalise Parker's work
\cite{Parker:1980:PRD} in allowing the reference curve $\gamma$ to be
accelerating, though the frame along it is still assumed to be
non-rotating (Fermi--Walker transported).  For vanishing acceleration
of $\gamma$, their result for the Dirac Hamiltonian coincides with
that of Parker.  Similar to Ito \cite{Ito:2021}, they consider a
`non-relativistic limit' by means of an expansion in ratios of
relevant energies to the rest energy, which effectively amounts to an
expansion in $1/m$.  Let it be mentioned already at this point that in
\cref{sec:comp_Perche_Neuser} we will show explicitly that the
expansion as presented in \cite{Perche.Neuser:2021} is \emph{not}
equivalent to the post-Newtonian expansion in $1/c$ that we employ.
This we believe, however, to be rooted in the expansion in
\cite{Perche.Neuser:2021} being inconsistently applied; when taking
proper care of all appearing terms, the expansion method of
\cite{Perche.Neuser:2021} is consistent with the corresponding
truncation of our results.

Our paper is an extension of those approaches, in that it also
includes inertial effects from acceleration and rotation to
consistently the same order as gravitational effects resulting from
curvature, namely to order $(\text{(geodesic distance)} /
\text{(charecteristic radii)})^2$.  We will find some inconsistencies
in the approximations of the aforementioned paper that result in the
omission of terms which we will restore.  Our paper is partly based on
the master's thesis \cite{Alibabaei:2022Thesis}.  Here we use the
opportunity to correct some oversights in the calculation of
order-$x^2$ terms in that thesis, that we will further comment on
below (cf.~\cref{fn:Ashkan_oversights}).

To sum up, our paper is organised as follows: In
\cref{sec:Dirac_eq_curv_spacet}, we recall the Dirac equation in
curved spacetime.  In \cref{sec:expansion_FNC}, we implement the first
step of our approximation procedure by expressing the Dirac equation
in generalised Fermi normal coordinates corresponding to an
accelerated reference worldline $\gamma$ and orthonormal, possibly
rotating frames along it.  In \cref{sec:expansion_post-Newtonian}, we
implement the second step, namely the `slow-velocity' post-Newtonian
expansion in $1/c$.  This step should be contrasted with the mentioned
$1/m$-expansions by others or expansions relying on Foldy--Wouthuysen
transformations.  In particular, this includes a comparison of our
resulting Hamiltonian to that obtained in \cite{Perche.Neuser:2021},
which we discuss in some detail in \cref{sec:comp_Perche_Neuser},
where we argue for an inconsistency within the calculation in
\cite{Perche.Neuser:2021}.  We conclude in \cref{sec:conclusion}.
Details of calculations and lengthy expressions are collected in
\cref{sec:app_conn_FN,sec:app_stat_geom,sec:app_Dirac_higher_order,sec:app_details_PN}.

\section{The Dirac equation in curved spacetime}
\label{sec:Dirac_eq_curv_spacet}

We consider a massive spin-half field $\psi$ in a general curved
background spacetime\footnote{Of course, for the very notion of spinor
  fields to make sense, we need to assume the spacetime to be equipped
  with a spin structure, i.e.\ a double cover of its orthonormal frame
  bundle such that the covering homomorphism is in trivialisations
  given by the double covering of the (homogeneous) Lorentz group
  $\mathcal L^\uparrow_+ = \mathrm{SO}_0(1,3)$ by the spin group
  $\mathrm{Spin}(1,3) = \mathrm{SL}(2,\mathbb C)$.  Dirac spinor
  fields are then sections of the Dirac spinor bundle, which is the
  vector bundle associated to the spin structure with respect to the
  $(\frac{1}{2},0) \oplus (0,\frac{1}{2})$ representation of the spin
  group.  As is well-known, the existence of a spin structure is for
  four-dimensional non-compact spacetimes equivalent to the spacetime
  manifold being parallelisable \cite{Geroch:1968}.}  $(M,g)$, coupled
to background electromagnetism, as described by the minimally coupled
Dirac equation
\begin{equation} \label{eq:Dirac}
  \Big(\I \gamma^I (\e_I)^\mu (\nabla_\mu - \I q A_\mu) - mc\Big) \psi
  = 0.
\end{equation}
Here $A_\mu$ are the components of the electromagnetic four-potential,
$\nabla$ is the Levi-Civita covariant derivative of the spacetime
metric $g$, extended to Dirac spinor fields, $m$ is the mass of the
field, and $q$ is its electric charge.  Note that we set $\hbar = 1$,
but keep explicit the velocity of light $c$.  A detailed exposition of
the Dirac equation in curved spacetime may be found in
\cite{Collas.Klein:2019}, to which we refer for further background
information.

The Dirac equation takes the above local form with respect to a choice
of \emph{tetrad}  $(\e_I) = (\e_0, \e_i)$, i.e.\ a local orthonormal
frame of vector fields.  Explicitly, this means that the vector fields
satisfy
\begin{equation}
  g(\e_I,\e_J) = \eta_{IJ}
\end{equation}
where $(\eta_{IJ}) = \mathrm{diag}(-1,1,1,1)$ are the components of
the Minkowski metric in Lorentzian coordinates.  The gamma matrices
$\gamma^I$ appearing in the Dirac equation \eqref{eq:Dirac} are the
standard Minkowski-spacetime gamma matrices $\gamma^I \in
\mathrm{End}(\C^4)$, which satisfy the Clifford algebra relation
\begin{align}
  \{\gamma^I ,\gamma^J \} = -2 \eta^{IJ} \mathbb{1}_4
\end{align}
with $\{\cdot,\cdot\}$ denoting the anti-commutator.  The Dirac
representation of the Lorentz algebra $\mathrm{Lie}(\mathrm{SO}(1,3))$
on $\C^4$ is given by
\begin{subequations}
\begin{align}
  \mathrm{Lie}(\mathrm{SO}(1,3)) \ni (\tensor{X}{^I_J})
  &\mapsto -\frac{1}{2} X_{IJ} S^{IJ} \in \mathrm{End}(\C^4), \\
  \intertext{with the generators $S^{IJ} \in \mathrm{End}(\C^4)$
  given by}
  S^{IJ} &= \frac{1}{4} [\gamma^I,\gamma^J].
\end{align}
\end{subequations}
Thus, the spinor covariant derivative is represented with respect to
the chosen tetrad by
\begin{subequations}
\begin{align}
  \nabla_\mu \psi &= \partial_\mu \psi + \Gamma_\mu \cdot \psi, \\
  \intertext{with the spinor representation of the local connection
  form explicitly given by}
  \Gamma_\mu &= -\frac{1}{2} \omega_{\mu IJ} S^{IJ}
\end{align}
\end{subequations}
in terms of the local connection form $\tensor{\omega}{_\mu^I_J}$ of
the Levi-Civita connection with respect to the tetrad, defined by
\begin{subequations} \label{eq:local_conn_form}
\begin{align}
  \nabla\e_I &= \tensor{\omega}{^J_I} \otimes \e_J \; , \\
  \intertext{i.e.\ in components}
  \nabla_\mu (\e_I)^\nu &= \tensor{\omega}{_\mu^J_I} (\e_J)^\nu \; .
\end{align}
\end{subequations}
Due to the local connection form taking values in the Lorentz algebra,
i.e.\ satisfying $\omega_{\mu IJ} = - \omega_{\mu JI}$, the spinor
representation of the connection form may be explicitly expressed as
\begin{equation} \label{eq:local_conn_form_spinor}
  \Gamma_\mu = -\frac{1}{2} \omega_{\mu IJ} S^{IJ}
  = -\frac{1}{2} \omega_{\mu 0i} \gamma^0 \gamma^i
    - \frac{1}{4} \omega_{\mu ij} \gamma^i \gamma^j .
\end{equation}

As said in the introduction, in the following we will describe a
systematic approximation scheme for the one-particle sector of the
massive Dirac theory from the point of view of an observer moving
along a fixed timelike reference worldline $\gamma$, which will
proceed in two conceptually independent steps.  The first step, which
is described in \cref{sec:expansion_FNC} and implements a
`weak-gravity' approximation by expanding the Dirac equation in
(generalised) Fermi normal coordinates, is actually valid without
restricting to the one-particle theory.

Only for the second step, the `slow-velocity' post-Newtonian expansion
in \cref{sec:expansion_post-Newtonian}, we will restrict to the
one-particle theory.  For this, we assume the spacetime and the
reference worldline $\gamma$ to be (approximately) stationary, such
that there is a well-defined (approximate) notion of particles in
quantum field theory and we may meaningfully restrict to the
one-particle sector of the theory.  This sector is then effectively
described by positive-frequency \emph{classical} solutions of the
Dirac equation, which we will approximate by the post-Newtonian
expansion.

\section{`Weak-gravity' expansion in generalised Fermi normal
  coordinates}
\label{sec:expansion_FNC}

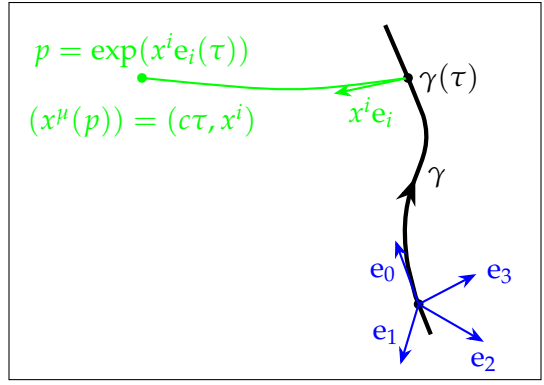
\begin{figure}
  \begin{center}
  \begin{tikzpicture}[thick]
    \draw[thin] (-3.25,-3) rectangle (3.75,2);
      
    \coordinate (A) at (2,1);
    \coordinate (B) at (-1.5,1);
    \draw[black,fill] (A) circle (.05) node[right] {$\gamma(\tau)$};
    \draw[green] (A) .. controls +(-1.5,-.2) .. (B);
    \draw[green,fill] (B) circle (.05) node[above]{$p
      = \exp(x^i \e_i(\tau))$};
    \node[green,below=.2] at (B) {$(x^\mu(p)) = (c \tau, x^i)$};
    \draw[green,-Stealth] (A) -- node[below] {$x^i \e_i$} +(192:1);

    \begin{scope}[ultra thick, decoration={markings, mark=at
        position 0.5 with {\arrow{Stealth}}}]
      \draw[postaction={decorate}] (2.3,-2.4) -- (2.1,-1.9) --
        (2,-1.5) .. controls (1.8,-.3) and (2.4,-.2)
        .. node[right] {$\gamma$} (2.2,.5) -- (2,1) -- (1.7,1.7);
    \end{scope}
    \filldraw [black] (2.14,-2) circle (.05);
    \draw[blue,-Stealth] (2.14,-2) -- node[left] {$\e_0$}
      +(110:.9);
    \draw[blue,-Stealth] (2.14,-2) -- node[left] {$\e_1$}
      (1.9,-2.8);
    \draw[blue,-Stealth] (2.14,-2) -- (3,-2.5)
      node[below] {$\e_2$};
    \draw[blue,-Stealth] (2.14,-2) -- (2.9,-1.6)
      node[right] {$\e_3$};
  \end{tikzpicture}
  \end{center}

  \caption{The construction of generalised Fermi normal coordinates}
  \label{fig:FNC}
\end{figure}

As the first step of our scheme, we will implement a `weak-gravity'
approximation of the Dirac equation with respect to a timelike
reference worldline $\gamma$ and orthonormal spacelike vector fields
$(\e_i(\tau))$ defined along $\gamma$ which are orthogonal to the
tangent $\e_0(\tau) := c^{-1} \dot\gamma(\tau)$.  The approximation
works by expressing the Dirac equation in \emph{generalised Fermi
  normal coordinates} with respect to $\gamma$ and $(\e_i)$.  These
coordinates are constructed as follows (compare \cref{fig:FNC}): in a
neighbourhood of $\gamma$, each point $p$ is connected to $\gamma$ by
a unique spacelike geodesic.  The temporal coordinate of $p$ is the
proper time parameter $\tau$ of the starting point of this geodesic,
defined with respect to some fixed reference point on $\gamma$, and
the spatial coordinates of $p$ are the components $x^i$ of the initial
direction of the geodesic with respect to the basis $(\e_i(\tau))$.
Phrased in terms of the exponential map, this means that the
coordinate functions $(x^\mu) = (c\tau, x^i)$ are defined by the
implicit equation
\begin{equation}
  p = \exp\left(x^i(p) \, \e_i(\tau(p))\right).
\end{equation}

These coordinates are adapted to an observer along $\gamma$ who
defines `spatial directions' using the basis $(\e_i)$.  Note that
differently to classical Fermi normal coordinates
\cite{Manasse.Misner:1963} we allow for the worldline $\gamma$ to
be accelerated---i.e.\ $\gamma$ need not be a geodesic---, as well as
for the basis $(\e_i)$ to be rotating with respect to
gyroscopes---i.e.\ the $(\e_i)$ need not be Fermi--Walker transported
along $\gamma$.  Generalised Fermi normal coordinates may be seen as
the best analogue of inertial coordinates that exists for an
arbitrarily moving observer carrying an arbitrarily rotating basis in
a general curved spacetime.

The acceleration $a(\tau)$ of $\gamma$ is the covariant derivative of
$\dot\gamma(\tau)$ along $\gamma$, i.e.\ the vector field
\begin{subequations}
\begin{align}
  a(\tau) &= \nabla_{\dot\gamma(\tau)} \dot\gamma(\tau) \\
  \intertext{along $\gamma$, which is everywhere orthogonal to
  $\dot\gamma = c \e_0$.  Note that we take the covariant derivative
  with respect to the worldline's four-velocity $\dot\gamma(\tau)$,
  such that the physical dimension of the components $a^\mu$ will
  really be that of an acceleration (given that the coordinate
  functions have the dimension of length).  The angular velocity of
  the observer's spatial basis vector fields $(\e_i)$ (with respect to
  non-rotating directions, i.e.\ Fermi--Walker transported ones) is
  another vector field along $\gamma$ that is everywhere orthogonal to
  $\dot\gamma = c \e_0$; we denote it by $\omega(\tau)$.  It is
  defined by}
  (\nabla_{\dot\gamma} \e_I)^\mu
          &= - (c^{-2} a^\mu \dot\gamma_\nu
            - c^{-2} \dot\gamma^\mu a_\nu
            + c^{-1} \tensor{\varepsilon}{_{\rho\sigma}^\mu_\nu}
              \dot\gamma^\rho \omega^\sigma) \,
            \e_I^\nu \; ,
\end{align}
\end{subequations}
where both sides of the equation are evaluated along $\gamma$, and
$\varepsilon$ denotes the volume form of the spacetime metric $g$.
The covariant derivatives of $a$ and $\omega$ along $\gamma$ will be
denoted by
\begin{equation}
  b(\tau) := \nabla_{\dot\gamma(\tau)} a(\tau), \quad
  \eta(\tau) := \nabla_{\dot\gamma(\tau)} \omega(\tau).
\end{equation}

When working in generalised Fermi normal coordinates we will denote
the timelike coordinate which has the dimension of length by $s = c
\tau$, since it is an extension of the proper length function along
$\gamma$.  In index notation, we will use $s$ as the timelike
coordinate index and reserve $0$ for use as the timelike index for
orthonormal frame components.

The components of the spacetime metric $g$ in generalised Fermi normal
coordinates may be expressed as formal power series in the geodesic
distance to $\gamma$ according to \cite{Ni.Zimmermann:1978}%
\begin{subequations} \label{eq:metric_FNC}
\begin{align}
  g_{ss} &= - 1 - 2 c^{-2} \vect a \cdot \vect x
           - c^{-4} (\vect a \cdot \vect x)^2 - R_{0l0m} x^l x^m
           + c^{-2} (\vect\omega \times \vect x)^2
           + \Or(\|\vect x\|^3), \\
  g_{si} &= c^{-1} (\vect\omega \times \vect x)_i
           - \frac{2}{3} R_{0lim} x^l x^m + \Or(\|\vect x\|^3), \\
  g_{ij} &= \delta_{ij} - \frac{1}{3} R_{iljm} x^l x^m
           + \Or(\|\vect x\|^3).
\end{align}
\end{subequations}
Here, in addition to the acceleration $a^i(\tau)$ of $\gamma$ and the
angular velocity $\omega^i(\tau)$ of the spatial basis $(\e_i)$, the
curvature tensor $R_{IJKL}(\tau)$ evaluated along $\gamma$ appears as
well; the components are taken with respect to the orthonormal basis
$(\e_0, \e_i)$ along $\gamma$.  We also have used standard
`three-vector' notation for geometric operations taking place in the
three-dimensional vector space $\Sigma_\tau = (\e_0(\tau))^\perp =
\mathrm{span}\{\e_i(\tau)\} \subset T_{\gamma(\tau)}M$ of the
observer's local `spatial directions', endowed with the Euclidean
metric $\delta_\tau := g|_{\Sigma_\tau}$ induced by $g$: we write
\begin{equation}
  \vect v \cdot \vect w := \delta_{ij} v^i w^j, \quad
  \|\vect v\| := \sqrt{\delta_{ij} v^i v^j}, \quad
  (\vect v \times \vect w)_i := \varepsilon_{ijk} v^j w^k
\end{equation}
for the scalar product, the norm, and the vector product with respect
to this metric.  Note that with respect to the orthonormal basis
$(\e_i)$, the components $\delta_{ij}$ of the induced metric and
$\varepsilon_{ijk}$ of its volume form are just given by the Kronecker
delta and the totally antisymmetric three-dimensional Levi-Civita
symbol, respectively.

The expansion in powers of the geodesic distance to $\gamma$
implements the desired approximation in terms of `weak gravity' and
`weak inertial effects': we expand according to
\begin{equation} \label{eq:FNC_small_params}
  R_{IJKL} \cdot \|\vect x\|^2 \ll 1, \qquad
  \frac{\vect a}{c^2} \cdot \vect x \ll 1, \qquad
  \frac{\vect\omega}{c} \cdot \vect x \ll 1, \qquad
  \frac{R_{IJKL;M}}{R_{NOPQ}} \cdot \|\vect x\| \ll 1,
\end{equation}
i.e.\ for the expansion to be valid at a point, the geodesic distance
to $\gamma$ has to be small compared to the curvature radius of
spacetime, the `acceleration radius' of $\gamma$, the `angular
velocity radius' of the spatial reference vector fields, and the
characteristic length scale on which the curvature changes.  This also
gives a precise analytical meaning to the formal expansion in the
dimensionful parameter $\|\vect x\|$: the actual dimensionless
quantity in which we expand is the ratio of $\|\vect x\|$ to the
minimum of the characteristic geometric lengths defined by the
spacetime curvature, acceleration $\vect a$, angular velocity
$\vect\omega$, and rate of change of the curvature, as given in
\eqref{eq:FNC_small_params}.  For the sake of brevity, in the
following we will speak of terms of $n^\text{th}$ order in the
geodesic distance to $\gamma$ simply as being of `order $x^n$', and
correspondingly use the shorthand notation $\Or(x^n) := \Or(\|\vect
x\|^n)$.

Our goal is to expand the Dirac equation \eqref{eq:Dirac}
systematically to order $x^2$.  To make precise what we mean by this,
first recall that for the local formulation \eqref{eq:Dirac} of the
Dirac equation to be possible, we have to choose a tetrad $(\e_I)$ not
only along the reference worldline $\gamma$, but also away from it.
This choice of tetrad is an additional input into the approximation
procedure, on top of the choice of local coordinate system.  However,
in our situation there is a natural choice for the tetrad: on
$\gamma$, we choose it to be given by the basis $(c^{-1} \dot\gamma,
\e_i)$ with respect to which the generalised Fermi normal coordinates
are defined; away from $\gamma$, we extend the vector fields by
parallel transport along spacelike geodesics.  The explicit form of
the tetrad components in coordinates will be computed at a later
stage.  With a choice of tetrad, we may rewrite the Dirac equation
\eqref{eq:Dirac} in the Schrödinger-like form
\begin{subequations} \label{eq:Dirac_Schrödinger}
\begin{align}
  \I \partial_\tau \psi
  &= H_\text{Dirac} \psi\\
  \intertext{with the Dirac Hamiltonian}
  \label{eq:Dirac_Hamiltonian_general}
  H_\text{Dirac}
  &= (g^{ss})^{-1} \gamma^J (\e_J)^s
      \Big(\I \gamma^I (\e_I)^i c (D_i + \Gamma_i) - m c^2\Big)
    - \I c (\Gamma_s - \I q A_s),
\end{align}
\end{subequations}
where we used that $\partial_s = c^{-1} \partial_\tau$ and that
$(g^{ss})^{-1} \gamma^J (\e_J)^s = \left(-\gamma^J (\e_J)^s
\right)^{-1}$, and where $D_i = \partial_i - \I q A_i$ denotes the
spatial electromagnetic covariant derivative.  It is \emph{this} Dirac
Hamiltonian that we will expand to order $x^2$ in the following.  Note
that the partial derivative $\partial_i = \frac{\partial}{\partial
  x^i}$ in the operator $D_i$ effectively is of order $x^{-1}$ when
acting on functions, such that in the following calculation, it is
important to keep track of terms of the form $x^l x^m x^n D_i$, which
despite their superficial appearance are in fact of order $x^2$.

We are now going to compute all objects appearing in the Dirac
Hamiltonian \eqref{eq:Dirac_Hamiltonian_general} to those orders in
$x$ which are necessary to obtain the total Hamiltonian to order
$x^2$.

In order to be able to expand covariant derivatives and the local
connection form to order $x^2$, we need to know the Christoffel
symbols in our coordinate system to order $x^2$.  Note that these
cannot be obtained from the metric components as given in
\eqref{eq:metric_FNC}: there the metric is given to order $x^2$, such
that its derivatives can only be known to order $x^1$.  However,
extending the work in \cite{Ni.Zimmermann:1978}, the Christoffel
symbols to order $x^2$ (and the metric to order $x^3$) in generalised
Fermi normal coordinates were calculated in \cite{Li.Ni:1979}.  The
Christoffel symbols are given in the appendix in
\eqref{eq:Christoffel_FNC} (note that some calculational errors were
made in \cite{Li.Ni:1979}, which we corrected in
\eqref{eq:Christoffel_FNC}).

We may now compute the coordinate components of our tetrad $(\e_I)$.
Recall that we define the tetrad by extending the vector fields
$(c^{-1} \dot\gamma, \e_i)$ along $\gamma$ into a neighbourhood of
$\gamma$ by parallel transport along spacelike geodesics.  Since
spacelike geodesics take a simple form in generalised Fermi normal
coordinates, the parallel transport equation may explicitly be solved
perturbatively using the Christoffel symbols
\eqref{eq:Christoffel_FNC}.  This calculation is straightforward, but
quite lengthy; it yields the tetrad components
\begin{subequations} \label{eq:frame}
\begin{align}
  (\e_0)^s &= 1 - c^{-2} \vect a \cdot \vect x
             + c^{-4} (\vect a \cdot \vect x)^2
             - \frac{1}{2} R_{0l0m} x^l x^m
             - \frac{1}{6} R_{0l0m;n} x^l x^m x^n \nonumber\\
           &\quad+ \frac{5}{6} c^{-2} (\vect a \cdot \vect x)
               R_{0l0m} x^l x^m
             - c^{-6} (\vect a \cdot \vect x)^3 + \Or(x^4),\\
  (\e_0)^i &= - c^{-1} (\vect\omega \times \vect x)^i
             + c^{-3} (\vect a \cdot \vect x)
               (\vect\omega \times\vect x)^i
             + \frac{1}{2} \tensor{R}{_{0l}^i_m} x^l x^m
             + \frac{1}{6} \tensor{R}{_{0l}^i_{m;n}} x^l x^m x^n
             \nonumber\\
           &\quad + \frac{1}{2} c^{-1} (\vect\omega \times \vect x)^i
               R_{0l0m} x^l x^m
             - c^{-5} (\vect\omega \times \vect x)^i
               (\vect a \cdot \vect x)^2
             - \frac{1}{3} c^{-2} (\vect a \cdot \vect x)
               \tensor{R}{_{0l}^i_m} x^l x^m \nonumber\\
           &\quad+ \Or(x^4),\\
  (\e_i)^s &= - \frac{1}{6} R_{0lim} x^l x^m
             - \frac{1}{12} R_{0lim;n} x^l x^m x^n
             + \frac{1}{6} c^{-2} (\vect a \cdot \vect x) R_{0lim}
               x^l x^m
             + \Or(x^4),\\
  (\e_i)^j &= \delta^j_i + \frac{1}{6} \tensor{R}{^j_{lim}} x^l x^m
             + \frac{1}{12} \tensor{R}{^j_{lim;n}} x^l x^m x^n
             + \frac{1}{6} c^{-1} (\vect\omega \times \vect x)^j
               R_{0lim} x^l x^m
             + \Or(x^4).
\end{align}
\end{subequations}
From this, we may compute the components of the dual frame as
\begin{subequations}
\begin{align}
  (\e^0)_s &= 1 + c^{-2} \vect a \cdot \vect x
             + \frac{1}{2} R_{0l0m} x^l x^m + \Or(x^3),\\
  (\e^0)_i &= \frac{1}{6} R_{0lim} x^l x^m + \Or(x^3),\\
  (\e^i)_s &= c^{-1} (\vect\omega \times \vect x)^i
             - \frac{1}{2} \tensor{R}{^i_{l0m}} x^l x^m + \Or(x^3),
             \displaybreak[0]\\
  (\e^i)_j &= \delta^i_j
             - \frac{1}{6} \tensor{R}{^i_{ljm}} x^l x^m + \Or(x^3).
\end{align}
\end{subequations}
Note that we have computed the dual frame components only to order
$x^2$ (instead of going to order $x^3$ as would have been possible
from \eqref{eq:frame}), since this suffices for our goal, namely the
expansion of the Dirac Hamiltonian
\eqref{eq:Dirac_Hamiltonian_general} to order $x^2$.

Now we have the required information in order to calculate the local
connection form $\tensor{\omega}{_\mu^I_J}$ according to
\eqref{eq:local_conn_form} to order $x^2$, which is given in the
appendix in \eqref{eq:local_conn_form_FNC}.  From this, we can
directly obtain its spinor representation $\Gamma_\mu$ according to
\eqref{eq:local_conn_form_spinor}.

We will also need the component $g^{ss}$ of the inverse metric to
order $x^3$.  Using the frame \eqref{eq:frame}, we may easily compute
this according to $g^{ss} = -((\e_0)^s)^2 + \delta^{ij} (\e_i)^s
(\e_j)^s$, yielding
\begin{align}
  g^{ss} &= -1 + 2 c^{-2} \vect a \cdot \vect x
           - 3 c^{-4} (\vect a \cdot \vect x)^2
           + 4 c^{-6} (\vect a\cdot \vect x)^3 \nonumber\\
         &\quad+ R_{0l0m} x^l x^m
           + \frac{1}{3} R_{0l0m;n} x^l x^m x^n
           - \frac{8}{3} c^{-2} (\vect a \cdot \vect x) R_{0l0m} x^l
             x^m
           + \Or(x^4).
\end{align}

We thus have obtained all ingredients to express the Dirac equation
\eqref{eq:Dirac}, \eqref{eq:Dirac_Schrödinger} in generalised Fermi
normal coordinates and our chosen tetrad to order $x^2$.  Inserting
$g^{ss}$, the tetrad components, and the spinor representation of the
local connection form as computed above into the Dirac Hamiltonian
\eqref{eq:Dirac_Hamiltonian_general}, by a tedious but straightforward
calculation, employing standard identities for products of three gamma
matrices, we obtain the explicit form of the Dirac Hamiltonian as
\begin{align} \label{eq:Dirac_FNC}
  H_\text{Dirac}
  &= \gamma^0 \left\{mc^2 + m \vect a \cdot \vect x
      + \frac{mc^2}{2} R_{0l0m} x^l x^m \right\}
      - \gamma^i \left\{\frac{mc^2}{6} R_{0lim} x^l x^m \right\}
    \nonumber\\
  &\quad + \mathbb{1} \bigg{\{}{-} q A_\tau
      + \I(\vect\omega \times \vect x)^i D_i
      - \frac{\I c}{2} \tensor{R}{_{0l}^i_m} x^l x^m D_i
      + \frac{\I c^{-1}}{4} (\vect a \cdot \vect x) R_{0l} x^l
        + \frac{\I c}{12} R_{0l;m} x^l x^m \nonumber\\
  &\qquad - \frac{\I c}{6} \tensor{R}{_{0l}^i_{m;n}} x^l x^m x^n D_i
      - \frac{\I c^{-1}}{6} (\vect a \cdot \vect x)
        \tensor{R}{_{0l}^i_m} x^l x^m D_i
    \bigg\} \nonumber\\
  &\quad - \gamma^0 \gamma^j \bigg\{\I c D_j
      + \frac{\I c^{-1}}{2} a_j
      + \I c^{-1}(\vect a \cdot \vect x) D_j
      + \frac{\I c}{4} (R_{0j0l} - R_{jl}) x^l
      + \frac{\I c}{2} R_{0l0m} x^l x^m D_j \nonumber\\
  &\qquad + \frac{\I c}{6} \tensor{R}{^i_{ljm}} x^l x^m D_i
      + \frac{\I c}{12} (R_{0j0l;m} - 2 R_{jl;m}) x^l x^m
      - \frac{\I c^{-1}}{4} (\vect a \cdot \vect x) R_{jl} x^l
    \nonumber\\
  &\qquad + \frac{\I c}{6} R_{0l0m;n} x^l x^m x^n D_j
      + \frac{\I c}{12} \tensor{R}{^i_{ljm;n}} x^l x^m x^n D_i
      + \frac{\I c^{-1}}{6} (\vect a \cdot \vect x) R_{0l0m} x^l x^m
        D_j \nonumber\\
  &\qquad + \frac{\I c^{-1}}{6} (\vect a \cdot \vect x)
        \tensor{R}{^i_{ljm}} x^l x^m D_i \bigg\} \nonumber\\
  &\quad + \gamma^i \gamma^j \bigg{\{}{-}\frac{\I}{4} \varepsilon_{ijk}
        \omega^k
      + \frac{\I c}{4} R_{0ijl} x^l
      + \frac{\I c}{6} R_{0lim} x^l x^m D_j
      + \frac{\I c}{12} R_{0ijl;m} x^l x^m \nonumber\\
  &\qquad + \frac{\I c}{12} R_{0lim;n} x^l x^m x^n D_j
      + \frac{\I c^{-1}}{6} (\vect a \cdot \vect x) R_{0lim} x^l x^m
        D_j
    \bigg\}
    + \Or(x^{3}).
\end{align}
As already stated in the introduction, this is our first main result.
Here $A_\tau = c A_s$ is the electric scalar potential with respect to
our coordinates.  Recall that the partial derivative operator
$\partial_i$ appearing in $D_i$ is effectively of order $x^{-1}$ when
acting on functions, such we need to keep terms of the form $x^l x^m
x^n D_i$ (since they are of order $x^2$).  The terms in the
Hamiltonian are ordered, in each pair of curly brackets, by order in
spatial geodesic distance $x$ to the worldline, with those terms of a
given order that include a $D_i$ appearing after those
without.\footnote{\label{fn:Ashkan_oversights}Here we correct some
  omissions that occurred in the master's thesis
  \cite{Alibabaei:2022Thesis} concerning terms of order $x^2$.
  Consequently our Dirac Hamiltonian \eqref{eq:Dirac_FNC} differs from
  that in \cite{Alibabaei:2022Thesis}.}

Note that setting $\omega = 0$ and ignoring quadratic terms in $a^i$
and $R_{IJKL}$ as well as terms involving covariant derivatives of the
curvature tensor, our Dirac Hamiltonian \eqref{eq:Dirac_FNC}
reproduces the Dirac Hamiltonian from \cite{Perche.Neuser:2021}.

\section{Post-Newtonian expansion}
\label{sec:expansion_post-Newtonian}

As the second step of our approximation scheme, we will now perform a
post-Newtonian `slow-velocity' expansion of the Dirac equation with
respect to our reference worldline $\gamma$.  In order to perform the
post-Newtonian expansion systematically, we are going to implement it
as a formal power series expansion\footnote{More precisely, since for
  some objects terms of negative order in $c^{-1}$ will appear, it is
  an expansion as formal \emph{Laurent} series.  We will however
  continue to use the term `power series', since most of our series
  will only have terms of non-negative order in $c^{-1}$.} in the
parameter $c^{-1}$, where $c$ is the velocity of light.\footnote{Of
  course, analytically speaking, a `Taylor expansion' in a
  dimensionful parameter like $c^{-1}$ does not make sense (even more
  so since $c$ is a constant of nature); only for \emph{dimensionless}
  parameters can a meaningful `small-parameter approximation' be made.
  In physical realisations of the limit from (locally) Poincaré- to
  Galilei-symmetric theories, this means that the corresponding small
  parameter has to be chosen as, e.g., the ratio of some typical
  velocity of the system under consideration to the speed of light.
  In the following, however, we will ignore such issues and simply
  expand in $c^{-1}$ as a formal `deformation' parameter.}  Such
formal expansions are a well-established device to implement Newtonian
limits and post-Newtonian expansions of (locally)
Poincaré-relativistic physics in a mathematically controlled manner:
they appear, of course, in the \.Inönü--Wigner contraction from the
Poincaré to the Galilei group \cite{Inonu.Wigner:1953}, and have been
applied, e.g., to systematically develop the post-Newtonian expansion
of the Klein--Gordon equation \cite{Giulini.Grossardt:2012,
  Schwartz.Giulini:2019:PNSE, Schwartz:2020,
  Giulini.Grossardt.Schwartz:2022}, or to discuss the rigorous
post-Newtonian expansion of General Relativity and its modifications
in the context of Newton--Cartan gravity (geometrised Newtonian
gravity) \cite{Dautcourt:1997,Tichy.Flanagan:2011,Hansen.EtAl:2019,
  Hansen.EtAl:2020,Hartong.EtAl:2023,Schwartz:2023}.  In order to
obtain a consistent post-Newtonian expansion\footnote{From a purely
  formal perspective, not assigning those $c^{-1}$-orders to the
  curvature components would lead to the expanded positive-frequency
  Dirac equation that we consider later not having perturbative
  solutions.  However, as already stated in the introduction, this
  assumption may also be viewed from a physical angle: in order for
  the acceleration of a system relative to $\gamma$, as given by the
  geodesic deviation equation, to stay bounded in the formal limit $c
  \to \infty$, we need to assume that $R_{0i0j} = \Or(c^{-2})$.}, we
need to treat the orthonormal-basis components of the curvature tensor
and its covariant derivative as being of order $c^{-2}$, i.e.
\begin{equation} \label{eq:curvature_c_order}
  R_{IJKL} = \Or(c^{-2}), R_{IJKL;M} = \Or(c^{-2}).
\end{equation}

Since we have already introduced a formal power series expansion in
$x$ (i.e.\ in spacelike geodesic distance to our reference worldline
$\gamma$), in the following we will encounter expressions that are
`doubly expanded' as power series in powers of both $c^{-1}$ and
$x$.\footnote{Formally, they will be valued in the formal
  Laurent/power series ring $\R((c^{-1},x]]$.}  When writing down
such expansions, we will order their terms as follows: first, we group
and sort the terms by order of $c^{-1}$, and second, the terms
comprising such a coefficient of a power $c^{-n}$ will be sorted by
order of $x$.  We will also use the notation $\Or(c^{-n}x^m)$ for
terms that are of order at least $n$ in the $c^{-1}$-expansion and
order at least $m$ in the $x$-expansion---e.g., we have $c^{-2}x^4 +
c^{-3}x^3 = \Or(c^{-2}x^3)$.  For example, the expansion of some
quantity $X$ might look like
\begin{equation} \label{eq:explain_sorting_double_exp}
  X = A + B_i x^i + C_{ij} x^i x^j + c^{-1} \left(E + F_i x^i \right)
  + \Or(c^{-1}x^2)
\end{equation}
(which would in particular imply that $X$ has \emph{vanishing}
coefficients for all powers $c^{-n}x$ with $n \ge 2$).

Considering the Dirac Hamiltonian $H_\text{Dirac}$ that appears in the
Dirac equation $\I \partial_\tau \psi = H_\text{Dirac} \psi$ in
generalised Fermi normal coordinates, as computed in
\eqref{eq:Dirac_FNC}, we may of course read off its expansion as a
power series in $c^{-1}$ directly from \eqref{eq:Dirac_FNC}---we just
need to keep in mind that we treat the curvature tensor as being of
order $c^{-2}$ according to \eqref{eq:curvature_c_order}.  However,
this expansion of the Dirac Hamiltonian in powers of $c^{-1}$ is of no
direct physical relevance for perturbation theory in the parameter
$c^{-1}$: from \eqref{eq:Dirac_FNC}, we directly obtain
$H_\text{Dirac} = \gamma^0 m c^2 + \Or(c^1)$, such that when expanding
the Dirac spinor field as a formal power series $\psi =
\sum_{k=0}^\infty c^{-k} \psi^{(k)}$, the Dirac equation tells us at
the lowest occurring order in $c^{-1}$, namely $c^2$, that $0 =
\gamma^0 m \psi^{(0)}$, i.e.\ $\psi^{(0)} = 0$.  At the next order
$c^1$, it then implies $\psi^{(1)} = 0$, etc.---meaning that the Dirac
equation has no non-trivial perturbative solutions of this form.
Hence, in order to obtain a meaningful `slow-velocity' approximation
to the Dirac theory, we need to make a different perturbative ansatz
for the spinor field.  This will be a \acronym{WKB}-like
`positive frequency' ansatz.

Conceptionally, we now restrict from the full Dirac quantum field
theory to its (effective) one-particle sector, which is a well-defined
notion if we assume the spacetime to be stationary.  The one-particle
sector is effectively described by \emph{classical} positive-frequency
solutions of the Dirac equation, where `positive frequency' is defined
with respect to the stationarity Killing field
\cite{Wald:1994}.\footnote{Often, this is called consideration of the
  `first-quantised theory'---a historically grown name that sometimes
  unfortunately tends to create conceptual confusion.  For details and
  caveats of how and why the one-particle sector of the quantum field
  theory is described by the positive-frequency classical theory, we
  refer to the extensive discussion in the monograph by Wald
  \cite{Wald:1994}.}  It is those positive-frequency solutions whose
field equation of motion we will expand in the following in powers of
$c^{-1}$.  A similar post-Newtonian expansion scheme for the
Klein--Gordon equation may be found in \cite{Giulini.Grossardt:2012,
  Schwartz.Giulini:2019:PNSE,Schwartz:2020}; a more general discussion
of such schemes is given in \cite{Giulini.Grossardt.Schwartz:2022}.

Note that in any realistic situation, in which the theory contains
interactions, this description can only be an approximation: the
energy of all processes taking place has to be small enough such as to
stay below the threshold of pair production, such that the system does
not leave the one-particle sector.  Therefore, such a post-Newtonian
expansion always has to be considered a low-energy approximation.

In the following, we will define positive frequencies with respect to
the coordinate time $\tau$ of the generalised Fermi normal coordinates
introduced in \cref{sec:expansion_FNC}; therefore, for the
relationship between positive-frequency classical solutions and the
one-particle sector of the quantum theory to (approximately) hold, we
need the timelike vector field $\partial/\partial\tau$ to be
(approximately) Killing.  The geometric meaning of this is briefly
discussed in \cref{sec:app_stat_geom}.  Note, however, that the
definition of positive-frequency solutions with respect to some `time
translation' vector field and the post-Newtonian expansion of such
solutions of course also works for time translation vector fields
which are \emph{not} Killing, i.e.\ in a non-stationary situation, in
which it still allows to view the full `relativistic'
positive-frequency Dirac equation as a formal deformation of its
(locally) Galilei-symmetric Newtonian limit.  In particular, as long
as we are in an \emph{approximately} stationary situation and the
vector field is \emph{approximately} Killing, the expansion will still
give an approximate description of the one-particle sector of quantum
field theory.

The \acronym{WKB}-like positive frequency ansatz that we will make for
the Dirac field will lead, due to the lowest $c^{-1}$ orders of the
Dirac equation, to a split of the Dirac spinor into two two-component
spinor fields with coupled equations of motion.  One of these
components can then, order by order in $c^{-1}$, be eliminated in
terms of the other, which will in the end lead to a Pauli equation for
the remaining two-spinor field, with gravitational and inertial
`corrections'.  We are going to carry out this expansion to order
$c^{-2}$, and in doing so, we want to keep the expansion in spacelike
geodesic distance to the reference worldline $\gamma$ such that the
resulting Pauli Hamiltonian contains terms to order $x^2$, as it was
the case for the Dirac Hamiltonian in \eqref{eq:Dirac_FNC}.  However,
in the decoupling/elimination process described above, the
to-be-eliminated component of the Dirac spinor field will be spatially
differentiated once.  Therefore, to achieve our goal of a consistent
expansion of the final Hamiltonian to order $x^2$, we actually need to
know those terms in the Dirac Hamiltonian which are of order up to
$c^{-1}$ in the $c^{-1}$-expansion \emph{not only to order $x^2$, but
  to order $x^3$}.  Employing the methods from \cite{Li.Ni:1979}, one
can calculate the order-$x^3$ terms in the Christoffel symbols in
generalised Fermi normal coordinates of $c^{-1}$-expansion order up to
$c^{-2}$ with a comparably small amount of work; and while doing so,
one can actually convince oneself that all $x$-dependent terms in the
Christoffel symbols are actually of order at least $c^{-2}$.  The
resulting Christoffel symbols, to order $x^3$ in the $c^{-2}$ terms
and to order $x^2$ in the higher-$c^{-1}$-order ones, are given in the
appendix in \eqref{eq:Christoffel_FNC_higher_order}.  Using these
further expanded Christoffel symbols, we can go through the further
steps of the calculation of the Dirac Hamiltonian from
\cref{sec:expansion_FNC}, thus computing the Dirac Hamiltonian to
order $x^3$ in the $c^{-1}$ terms and to order $x^2$ in the
higher-$c^{-1}$-order ones.  The expressions for the frame, the
connection form and the inverse metric component $g^{ss}$ arising as
intermediate results in this process are given in
\cref{sec:app_Dirac_higher_order}; the resulting Dirac Hamiltonian is
given in \eqref{eq:Dirac_FNC_higher_order}.  \emph{This} Dirac
Hamiltonian will give rise, when carrying out our systematic expansion
of the positive-frequency Dirac equation in powers of $c^{-1}$, to a
consistently derived Pauli Hamiltonian to order $x^2$ and $c^{-2}$.

As the first step for implementing the expansion, we make for the
Dirac field the \acronym{WKB}-like ansatz\footnote{Note that in the
  master's thesis \cite{Alibabaei:2022Thesis} on which the present
  article is based, a different notational convention was used in
  which $\tilde{\psi}^{(k)}$ includes the factor of $c^{-k}$.}
\begin{equation}
  \psi = \e^{\I c^2 S} \tilde{\psi} \; \text{with} \;
  S = \Or(c^0),
  \tilde{\psi} = \sum_{k=0}^\infty c^{-k} \tilde{\psi}^{(k)}.
\end{equation}
This ansatz we then insert into the Dirac equation $\I \partial_\tau
\psi = H_\text{Dirac} \psi$, with the Dirac Hamiltonian given by
\eqref{eq:Dirac_FNC_higher_order}.  The resulting equation we multiply
with $\e^{-\I c^2 S}$ and compare coefficients of different powers of
$c^{-1}$.  At the lowest ocurring order $c^3$, we obtain the equation
$0 = \gamma^0 \gamma^i (\partial_i S) \tilde{\psi}^{(0)}$, which in
order to allow for non-trivial solutions $\tilde{\psi}$ enforces
$\partial_i S = 0$, i.e.\ the function $S$ depends only on time.  At
the next order $c^2$, we then obtain the equation
\begin{equation}
  -(\partial_\tau S) \tilde{\psi}^{(0)}
  = \gamma^0 m \tilde{\psi}^{(0)}.
\end{equation}
Since $\gamma^0$ has eigenvalues $\pm1$, for non-trivial solutions
$\tilde{\psi}$ of the Dirac equation to exist we need $\partial_\tau S
= \pm m$.  Since we are interested in positive-frequency solutions of
the Dirac equation, we choose $S = -m\tau$, discarding the constant of
integration (which would lead to an irrelevant global phase).  The
preceding equation then tells us that the component of
$\tilde{\psi}^{(0)}$ which lies in the $-1$ eigenspace of $\gamma^0$
has to vanish.

In the following, we will work in the Dirac representation for the
gamma matrices, in which they are given by
\begin{equation}
  \gamma^0 = \begin{pmatrix} \mathbb{1} & 0 \\
    0 & -\mathbb{1}
  \end{pmatrix}, \quad
  \gamma^i = \begin{pmatrix} 0 & \sigma^i \\
    -\sigma^i & 0
  \end{pmatrix}
\end{equation}
in terms of the Pauli matrices $\sigma^i$, such that Dirac spinors may
be decomposed as
\begin{equation}
  \psi = \begin{pmatrix} \psi_A \\ \psi_B \end{pmatrix}
\end{equation}
in terms of their components $\psi_A,\psi_B$ lying in the $+1$ and
$-1$ eigenspace of $\gamma^0$, respectively.  (Note that $\psi_{A,B}$
are represented by functions taking values in $\C^2$.)

Summing up the above, our ansatz for the Dirac field now takes the
form
\begin{equation} \label{eq:Dirac_PN_ansatz}
  \psi = \e^{-\I mc^2 \tau}
  \begin{pmatrix}
    \tilde{\psi}_A \\ \tilde{\psi}_B
  \end{pmatrix}, \quad
  \tilde{\psi}_{A,B}
  = \sum_{k=0}^\infty c^{-k} \tilde{\psi}_{A,B}^{(k)} \; ,
\end{equation}
and we know that $\tilde{\psi}_B^{(0)} = 0$.  Inserting this into the
Dirac equation and multiplying with $\e^{\I mc^2 \tau}$, we obtain two
coupled equations for $\tilde{\psi}_{A,B}$, which are given in the
appendix in \eqref{eq:Dirac_expanded}.  Now comparing in these
equations the coefficients of different orders of $c^{-1}$, we may
order by order read off equations for the $\tilde{\psi}_{A,B}^{(k)}$.
These allow to eliminate $\tilde{\psi}_B$ in favour of
$\tilde{\psi}_A$, for which we will obtain a post-Newtonian Pauli
equation.

More explicitly, this proceeds as follows.
\eqref{eq:Dirac_expanded_1} at order $c^1$ yields 
\begin{align}
  0 = -\I \sigma^j D_j \tilde{\psi}_B^{(0)} \; ,
\end{align}
which is trivially satisfied since $\tilde{\psi}_B^{(0)} = 0$.
\eqref{eq:Dirac_expanded_2} at order $c^1$ gives
\begin{align} \label{eq:PN_elim_psi_B_1}
  2m \tilde{\psi}_B^{(1)} = -\I \sigma^j D_j \tilde{\psi}_A^{(0)}
\end{align}
and thus allows us to express $\tilde{\psi}_B^{(1)}$ in terms of
$\tilde{\psi}_A^{(0)}$.  We can carry on to the next order:
\eqref{eq:Dirac_expanded_1} at order $c^0$ yields
\begin{equation}
  \bigg\{\I \partial_\tau + q A_\tau - m \vect a \cdot \vect x
    - \frac{mc^2}{2} R_{0l0m}x^l x^m
    - \I (\vect\omega \times \vect x)^i D_i
    + \frac{1}{2} \vect\sigma \cdot \vect\omega
    + \Or(x^3)
    \bigg\} \tilde{\psi}_A^{(0)}
    = - \I \sigma^j D_j \tilde{\psi}_B^{(1)} .
\end{equation}
Using \eqref{eq:PN_elim_psi_B_1}, this may be rewritten as a Pauli
equation
\begin{equation} \label{eq:Pauli_PN_0}
  \I \partial_t \tilde{\psi}_A^{(0)} = H^{(0)} \tilde{\psi}_A^{(0)}
\end{equation}
for $\tilde{\psi}_A^{(0)}$, with lowest-order Hamiltonian
\begin{equation} \label{eq:Pauli_PN_Ham_0}
  H^{(0)} = -\frac{1}{2m} (\vect\sigma \cdot \vect D)^2
  + m \vect a \cdot \vect x
  + \frac{mc^2}{2} R_{0l0m} x^l x^m
  + \I (\vect\omega \times \vect x)^i D_i
  - \frac{1}{2} \vect\sigma \cdot \vect\omega
  - q A_\tau + \Or(x^3) .
\end{equation}

Next, \eqref{eq:Dirac_expanded_2} at order $c^0$ allows us to express
$\tilde{\psi}_B^{(2)}$ in terms of $\tilde{\psi}_A^{(1)}$ and
$\tilde{\psi}_A^{(0)}$:
\begin{align} \label{eq:PN_elim_psi_B_2}
  2m \tilde{\psi}_B^{(2)} = -\I \sigma^j D_j \tilde{\psi}_A^{(1)}
  + \left(\frac{mc^2}{6} \sigma^i R_{0lim} x^l x^m
    + \frac{m c^2}{12} \sigma^i R_{0lim;n} x^l x^m x^n
    + \Or(x^4) \right) \tilde{\psi}_A^{(0)}
\end{align}
Note that since $\tilde{\psi}_B^{(2)}$ will be differentiated once in
the following calculation, here we need to include the term of order
$x^3$ for later consistency, i.e.\ in order to be able to obtain the
final Hamiltonian to order $x^2$.  This is why we needed to know the
low-$c^{-1}$-order terms of the Dirac Hamiltonian to order $x^3$, and
not just order $x^2$.  The same will happen at several later stages of
the computation.

\eqref{eq:Dirac_expanded_1} at order $c^{-1}$ will then give an
equation for $\tilde{\psi}_A^{(1)}$, which may be rewritten in the
Pauli-like form
\begin{equation} \label{eq:Pauli_PN_1}
  \I \partial_t \tilde{\psi}_A^{(1)}
  = H^{(0)} \tilde{\psi}_A^{(1)} + H^{(1)} \tilde{\psi}_A^{(0)} \; .
\end{equation}
Due to the nature of the expansion, the lowest-order operator
$H^{(0)}$ read off here will be the same as the one from the previous
order.  Detailed expressions may be found in
\cref{sec:app_details_PN}.

Continuing, \eqref{eq:Dirac_expanded_2} at order $c^{-1}$ allows to
express $\tilde{\psi}_B^{(3)}$ in terms of $\tilde{\psi}_A^{(2)}$,
$\tilde{\psi}_A^{(1)}$, $\tilde{\psi}_A^{(0)}$, and
$\tilde{\psi}_B^{(1)}$, which in turn may be expressed in terms of
$\tilde{\psi}_A^{(0)}$ by \eqref{eq:PN_elim_psi_B_1}.
\eqref{eq:Dirac_expanded_1} at order $c^{-2}$ can then be rewritten as
the Pauli-like equation
\begin{equation} \label{eq:Pauli_PN_2}
  \I \partial_t \tilde{\psi}_A^{(2)}
  = H^{(0)} \tilde{\psi}_A^{(2)} + H^{(1)} \tilde{\psi}_A^{(1)}
  + H^{(2)} \tilde{\psi}_A^{(0)} \; .
\end{equation}
Again, we know that $H^{(0)}$ and $H^{(1)}$ are the same as determined
before; the operator $H^{(2)}$ will contain new information.  Detailed
expressions may again be found in \cref{sec:app_details_PN}.  Note
that in the process of expressing $\tilde{\psi}_B^{(3)}$ in terms of
the $\tilde{\psi}_A$, one term arises for which we need to re-use the
Pauli equation \eqref{eq:Pauli_PN_0} for $\tilde{\psi}_A^{(0)}$ in
order to fully eliminate the time derivative in the resulting
expression.

The three Pauli-like equations \eqref{eq:Pauli_PN_0},
\eqref{eq:Pauli_PN_1} and \eqref{eq:Pauli_PN_2} now may be combined
into a Pauli equation
\begin{subequations}
\begin{align}
  \I \partial_t \tilde{\psi}_A
  &= H_\text{Pauli} \tilde{\psi}_A \\
  \intertext{with Hamiltonian}
  H_\text{Pauli}
  &= H^{(0)} + c^{-1} H^{(1)} + c^{-2} H^{(2)} + \Or(c^{-3}).
\end{align}
\end{subequations}
Explicitly, the post-Newtonian Pauli Hamiltonian reads
\begin{align} \label{eq:Pauli_PN}
  H_\text{Pauli}
  &= \bigg\{ \textcolor{green}{{}-\frac{1}{2m}}
      - \frac{1}{2mc^2} \vect a \cdot \vect x
      - \frac{1}{4m} R_{0l0m} x^l x^m
      - \frac{1}{8m} R_{0l0m;n} x^l x^m x^n \nonumber\\
  &\qquad - \frac{1}{24m} R_{0k0l;mn} x^k x^l x^m x^n \bigg\}
      \textcolor{green}{{}(\vect\sigma \cdot \vect D)^2}
    \textcolor{green}{{}- \frac{1}{8m^3c^2}
      (\vect\sigma \cdot \vect D)^4} \nonumber\\
  &\quad + \left\{-\frac{1}{6m} \tensor{R}{^i_l^j_m} x^l x^m
      - \frac{1}{12m} \tensor{R}{^i_l^j_{m;n}} x^l x^m x^n
      - \frac{1}{40m} \tensor{R}{^i_k^j_{l;mn}} x^k x^l x^mx^n
    \right\} D_i D_j \nonumber\\
  &\quad + \bigg\{\I(\vect\omega \times \vect x)^j
      - \frac{2\I c}{3} \tensor{R}{_{0l}^j_m} x^l x^m
      - \frac{\I c}{4} \tensor{R}{_{0l}^j_{m;n}} x^l x^m x^n
      - \frac{1}{4mc^2} a^j
      - \frac{\I}{4mc^2} (\vect\sigma \times \vect a)^j \nonumber\\
  &\qquad + \frac{1}{12m} (4\tensor{R}{^j_l} + \tensor{R}{_0^j_{0l}})
        x^l
      + \frac{\I}{8m} \sigma^k
        (-2 \tensor{\varepsilon}{^{ij}_k} R_{0l0i}
        + \tensor{\varepsilon}{^{im}_k} \tensor{R}{^j_{lim}}) x^l \nonumber\\
  &\qquad + \frac{1}{24m}
      \Big(5 \tensor{R}{^j_{l;m}} - 3 \tensor{R}{_0^j_{0l;m}}
        - \tensor{R}{_{0l0m}^{;j}}
        - \tensor{R}{^j_l^i_{m;i}} \nonumber\\
  &\qquad\quad - \I \tensor{\varepsilon}{^{ij}_k} \sigma^k
          (2R_{0i0l;m} + R_{0l0m;i})
        + 2\I \tensor{\varepsilon}{^{in}_k} \sigma^k
          \tensor{R}{^j_{lin;m}} \Big) x^l x^m \nonumber\\
  &\qquad + \frac{1}{120m} \Big(9 \tensor{R}{^j_{l;mn}}
        - 6 \tensor{R}{_0^j_{0l;mn}}
        - 5 \tensor{R}{_{0l0m}^{;j}_n}
        - 3 \tensor{R}{^j_l^i_{m;in}}\Big)
      x^l x^m x^n \nonumber\\
  &\qquad + \frac{\I}{96m} \sigma^k \Big(
        {-4} \tensor{\varepsilon}{^{ij}_k} (R_{0i0l;mn} + R_{0l0m;ni})
        + 3 \tensor{\varepsilon}{^{ir}_k} \tensor{R}{^j_{lir;mn}}\Big)
      x^l x^m x^n \bigg\} D_j \nonumber\\
  &\quad \textcolor{green}{{}- q A_\tau
    + m \vect a \cdot \vect x
    + \frac{mc^2}{2} R_{0l0m} x^l x^m
    - \frac{1}{2} \vect\sigma \cdot \vect\omega} \nonumber\\
  &\quad + \frac{\I c}{3} R_{0l} x^l
    - \frac{c}{4} \tensor{\varepsilon}{^{ij}_k} \sigma^k R_{0lij}
      x^l
    + \frac{\I c}{24}(5 R_{0l;m} - \tensor{R}{_{0l}^i_{m;i}}) x^l x^m
    - \frac{c}{8} \tensor{\varepsilon}{^{ij}_k} \sigma^k R_{0lij;m}
      x^l x^m \nonumber\\
  &\quad + \frac{1}{8m} R + \frac{1}{4m} R_{00}
    + \frac{1}{16m} (R_{;l} + 2 \tensor{R}{^i_{l;i}}) x^l
    + \frac{\I}{24m} \tensor{\varepsilon}{^{ij}_k} \sigma^k
      (R_{0i0l;j} - 2 R_{il;j}) x^l \nonumber \displaybreak[0]\\
  &\quad + \frac{1}{48m} \Big(R_{;lm} + 4 \tensor{R}{^i_{l;im}}
        + \I \tensor{\varepsilon}{^{ij}_k} \sigma^k (R_{0i0l;jm}
          - 3 R_{il;jm})\Big)
      x^l x^m \nonumber\\
  &\quad - \frac{q}{4m^2c^2} \sigma^i \sigma^j D_i E_j
    - \frac{q}{12m} (R_{lm} + R_{0l0m}) x^l x^m
      \vect\sigma \cdot \vect B
    + \frac{q}{12m} \sigma^j R_{iljm} x^l x^m B^i \nonumber\\
  &\quad + \frac{\I q}{4m^2c^2} \vect\sigma \cdot
      (\vect\omega \times \vect B)
    + \frac{q}{2m^2c^2} \vect\omega \cdot \vect B
    + \frac{q}{4m^2c^2} (\omega_j x^i - \omega^i x_j)
      D_i B^j \nonumber\\
  &\quad + \frac{\I q}{4m^2c^2} (\vect\sigma \cdot
      (\vect\omega \times \vect x)) \vect B \cdot \vect D
    - \frac{\I q}{4m^2c^2} \sigma^j
      (\vect\omega \times \vect x) \cdot \vect D B_j
    + \Or(c^{-3}) + \Or(x^3),
\end{align}
where $E_i = \partial_i A_\tau - \partial_\tau A_i$ is the electric
field and $B^i = \varepsilon^{ijk} \partial_j A_k$ is the magnetic
field (note that up to higher-order corrections, these are indeed the
electromagnetic field components in an orthonormal basis).  Note that
in the expressions $D_i E_j$, $D_i B^j$, and $\vect D B_j$, the $D_i$
acts on the product of the electric/magnetic field and the
$\tilde{\psi}_A$ on which the Hamiltonian acts.  The post-Newtonian
Pauli Hamiltonian \eqref{eq:Pauli_PN} is the second main result of
this paper.

The terms in the Hamiltonian are ordered as follows: the terms
involving electromagnetic fields come in the end, the terms without in
the beginning.  The latter are grouped by the form of the spatial
derivative operators (built from $D_i$) appearing in them.  In each of
these groups, the terms are ordered as explained before
\eqref{eq:explain_sorting_double_exp}: first, they are sorted by order
of $c^{-1}$, and for each $c^{-1}$-order, the terms are sorted by
order of $x$.

The lowest-order terms in the Hamiltonian, marked in green in
\eqref{eq:Pauli_PN}, have clear interpretations: we have the usual
`Newtonian' kinetic-energy term $-\frac{1}{2m} (\vect\sigma \cdot
\vect D)^2$ for a Pauli particle minimally coupled to
electromagnetism, the coupling $-q A_\tau$ to the electric scalar
potential, the `Newtonian' gravitational coupling $m(\vect a \cdot
\vect x + \frac{c^2}{2} R_{0l0m} x^l x^m)$ to a potential including an
acceleration and a tidal force term, and the spin--rotation coupling
$-\frac{1}{2} \vect\sigma \cdot \vect\omega$.  Note also that the
Hamiltonian contains the special-relativistic correction to kinetic
energy, $-\frac{1}{8m^3c^2} (\vect\sigma \cdot \vect D)^4$.  The other
terms are higher-order inertial and gravitational corrections.

Note that the scalar product of our quantum theory, with respect to
which the Hamiltonian \eqref{eq:Pauli_PN} needs to be interpreted, is
\emph{not} simply the standard $\mathrm{L}^2$ scalar product of
$\mathbb C^2$-valued Pauli wavefunctions
\begin{equation}
  \langle\tilde{\phi}_A, \tilde{\psi}_A\rangle_{\mathrm{L}^2}
  := \int\D^3x \, \overline{\tilde{\phi}_A(\vect x)}^T
    \tilde{\psi}_A(\vect x).
\end{equation}
Rather, the correct scalar product is that coming from the original
Dirac theory: we start with the original Dirac scalar product
\begin{equation}
  \langle\phi, \psi\rangle_\text{Dirac}
  := \int_\Sigma \D\mathrm{vol}_\Sigma \, n_\mu \,
    \overline{\phi}^T \gamma^I (\e_I)^0
    \gamma^J (\e_J)^\mu \psi
\end{equation}
and compute its expansion in $x$ and $c^{-1}$ that arises from
inserting our post-Newtonian ansatz \eqref{eq:Dirac_PN_ansatz} for the
Dirac field and expressing $\tilde{\phi}_B$ and $\tilde{\psi}_B$ in
terms of $\tilde{\phi}_A$ and $\tilde{\psi}_A$.  With respect to
\emph{this} scalar product, the Hamiltonian is automatically
Hermitian, since the Dirac scalar product in the full theory is
conserved under time evolution.

Our post-Newtonian quantum theory also comes with a natural position
operator, which in this representation of the Hilbert space is given
by multiplication of `wave functions' by coordinate position $x^a$.
This operator arises as the post-Newtonian equivalent of that operator
in the one-particle sector of the full Dirac theory which multiplies
the Dirac fields by coordinate position $x^a$.  For the case of the
reference worldline $\gamma$ being an inertial worldline in Minkowski
spacetime and a non-rotating frame, that operator is, in fact, the
Newton--Wigner position operator \cite{Newton.Wigner:1949,
  Schwartz.Giulini:2020}.

\subsection{Comparison to previous results by others}
\label{sec:comp_Perche_Neuser}

We now want to compare our post-Newtonian Hamiltonian
\eqref{eq:Pauli_PN} to that obtained in \cite{Perche.Neuser:2021}.  In
order to do so we proceed as follows: we first recall the hypotheses
on which the expansion in \cite{Perche.Neuser:2021} is based.  These
we then use to further approximate our result in accordance with these
hypotheses.  Then, finally, we compare the result so obtained with
that of \cite{Perche.Neuser:2021}.  We shall find a difference which
we interpret as an inconsistency in~\cite{Perche.Neuser:2021}.

Now, the approximation hypotheses in \cite{Perche.Neuser:2021} that go
beyond those imposed by us fall into three classes: First, concerning
`weak gravity', they assume $\omega = 0$ (no frame rotation), they
neglect quadratic terms in $a^i$ and $R_{IJKL}$, and, finally, they
also do not consider terms involving covariant derivatives of the
curvature tensor.  Second, as regards their `non-relativistic
approximation', they neglect terms of quadratic or higher order in
$m^{-1}$.  Third, they trace over the spin degrees of freedom, i.e.\
compute $\frac{1}{2} \mathrm{tr}(H_\text{Pauli})$, in order to obtain
what in \cite[p.\,16]{Perche.Neuser:2021} was called `the Hamiltonian
[$\ldots$] compatible with the description of a Schrödinger
wavefunction'\footnote{This method of tracing over the spin degrees of
  freedom in order to obtain a Hamiltonian acting on single-component
  (complex-number-valued) wavefunctions is used in
  \cite{Perche.Neuser:2021} without further justification beyond the
  goal of acting on $\mathbb{C}$-valued functions.  We do not believe
  this method to be of general physical validity for the following
  reason: the unitary time evolution described by the full
  post-Newtonian Pauli Hamiltonian contains interactions between the
  position and spin degrees of freedom.  Therefore, the effective time
  evolution which we would obtain by ignoring the spin, i.e.\ by
  taking the partial trace of the total density matrix over the spin
  degrees of freedom, would no longer be unitary.  Consequently, it
  cannot be described by a Schrödinger equation with respect to some
  Hamiltonian.  Of course, this general argument does not exclude
  that, depending on the context, an approximately unitary time
  evolution for some specific initial states does indeed exist, but
  such an argument is not given in \cite{Perche.Neuser:2021}.
  Nevertheless, for the sake of comparison to
  \cite{Perche.Neuser:2021}, we still apply the tracing procedure
  which we consider physically unwarranted.}.  In units with $c = 1$,
as used in \cite{Perche.Neuser:2021}, the result of applying this
procedure to our Hamiltonian reads
\begin{align}
  \frac{1}{2} \mathrm{tr}(H_\text{Pauli})
    &= \bigg\{ -\frac{1}{2m} -\frac{1}{2m} \vect a \cdot \vect x
      - \frac{1}{4m} R_{0l0m} x^l x^m  \bigg\} \vect D^2 \nonumber\\
  &\quad + \bigg\{- \frac{2\I}{3} \tensor{R}{^j_{l0m}} x^l x^m
      - \frac{1}{4m} a^j
      + \frac{1}{3m} \tensor{R}{^j_l} x^l
      + \frac{1}{12m} \tensor{R}{_{0l0}^j} x^l \bigg\} D_j \nonumber\\
  &\quad - q A_\tau
    + m \vect a \cdot \vect x
    + \frac{m}{2} R_{0l0m} x^l x^m
    + \frac{\I}{3} R_{0l} x^l \nonumber\\
  &\quad + \frac{1}{8m} R + \frac{1}{4m} R_{00}
    - \frac{1}{6m} \tensor{R}{^i_l^j_m} x^l x^m D_j D_i \; .
\end{align}
This is \emph{different} from the resulting Hamiltonian $\mathcal H$
from \cite{Perche.Neuser:2021}, with the difference reading
\begin{align} \label{eq:diff_Perch_Neu}
  \frac{1}{2} \mathrm{tr}(H_\text{Pauli}) - \mathcal H
  &= \bigg\{ \frac{1}{4m} \vect a \cdot \vect x
      + \frac{1}{8m} R_{0l0m} x^l x^m \bigg\} \vect D^2
    + \bigg\{ \frac{1}{2m} a^j
      + \frac{1}{2m} \tensor{R}{_{0l0}^j} x^l \bigg\} D_j
    + \frac{1}{4m} R_{00} \; .
\end{align}
This difference arises precisely from that term in the computation of
our second-order Hamiltonian $H^{(2)}$ for which we had to re-use the
lowest-order Pauli equation for $\tilde{\psi}_A^{(0)}$: in the final
Pauli Hamiltonian, this term amounts to a contribution of
\begin{equation} \label{eq:diff_Perch_Neu_our_calc}
  - \frac{1}{4m^2c^2} (-\I \vect\sigma \cdot \vect D)
  \{ \I q \vect\sigma \cdot \vect E
    + (-\I \vect\sigma \cdot \vect D) (H^{(0)} + q A_\tau) \};
\end{equation}
due to $H^{(0)}$ containing terms proportional to $m$, this expression
contains terms proportional to $m^{-1}$, which yield exactly the
difference term \eqref{eq:diff_Perch_Neu}.

Closely examining the calculation of \cite{Perche.Neuser:2021}, one
can exactly pinpoint the step of this calculation at which the above
term has been neglected: in appendix C of \cite{Perche.Neuser:2021},
going from equation (C3) to (C4), an inverse operator of the form
$\frac{1}{2m} \big(1 + \frac{\I \partial_T + q A_0}{2m} + (\text{terms
  linear in $a^i$ and $R$})\big)^{-1}$ is evaluated via a perturbative
expansion (in the notation of \cite{Perche.Neuser:2021}, $\I
\partial_T$ is the `non-relativistic energy' operator, i.e.\ the total
energy of the Dirac solution minus the rest energy).  The authors of
\cite{Perche.Neuser:2021} argue that when expanding (with respect to
small quotients of involved energies), `the rest mass of the system
tends to be much larger than any of the terms that show up in the
expansion', such that `in a power expansion of the inverse operator in
equation (C3), it makes sense to neglect terms that will contribute
with order $m^{-2}$'.  Following this argument, the term involving
$\frac{\I\partial_T}{2m}$ is neglected.  However, by following the
ensuing calculation one can check that if it were \emph{not} neglected
at this point, this term would in the end lead to a contribution to
the final Pauli Hamiltonian of the form
\begin{equation}
  -\frac{1}{4m^2}(-\I \vect\sigma \cdot \vect D)^2 H^{(0)}
  + \Or(m^{-2})
\end{equation}
in our notation, which to the order of approximation used in
\cite{Perche.Neuser:2021} is precisely the term noted above in
\eqref{eq:diff_Perch_Neu_our_calc}.

We thus see that the $\frac{\I\partial_T}{2m}$ term ought \emph{not}
to be neglected in going from (C3) to (C4) in
\cite{Perche.Neuser:2021}, since in the end it leads to terms that are
\emph{of the same order} as the other correction terms.  A more direct
formulation of the argument against neglecting this term is to note
that $\I\partial_T$ acting on $\psi_A$ (in the notation of
\cite{Perche.Neuser:2021}) induces terms proportional to $m$, such
that $-\I\partial_T/(4m^2)$ is not actually of order $m^{-2}$, but of
order $m^{-1}$.

One may also formulate our argument against the neglection without
referring to expanding in $m^{-1}$ at all, speaking only about
quotients of energies instead, in the spirit of
\cite{Perche.Neuser:2021}: if one were to neglect the term
$-\I\partial_T/(4m^2) = -\frac{1}{2m} \cdot
\frac{\text{`non-relativistic energy'}}{2(\text{rest energy})}$ in
going from (C3) to (C4) in \cite{Perche.Neuser:2021}, then one would
as well have to neglect the terms $-\frac{1}{2m}(\frac{1}{2} a_j x^j +
\frac{1}{4} R_{k0m0} x^k x^m) = -\frac{1}{2m} \cdot \frac{m(a_j x^j +
  R_{k0m0} x^k x^m/2)}{2m} = -\frac{1}{2m} \cdot
\frac{\text{corrections in `non-relativistic energy'}}{2(\text{rest
    energy})}$.  These last terms, however, clearly have to be kept in
the calculation since they contribute at a relevant order, and indeed
are kept in \cite{Perche.Neuser:2021}.

Thus, we come to the conclusion that the difference between the result
of \cite{Perche.Neuser:2021} and our result when truncated to linear
approximation order is due to an undue neglection in
\cite{Perche.Neuser:2021}, which without further justification seems
to render the approximation used in \cite{Perche.Neuser:2021}
inconsistent.  In our opinion, this exemplifies that a mathematically
clear systematic approximation scheme with spelled-out
assumptions---such as ours, based on (formal) power series expansions
in deformation parameters---reduces possibilities for conceptual
errors in approximative calculations.

\section{Conclusion}
\label{sec:conclusion}

Deducing the impact of classical gravitational fields (in the sense of
General Relativity) onto the dynamical evolution of quantum systems is
a non-trivial task of rapidly increasing theoretical interest given
the acceleration that we currently witness in experimental areas, like
$g$-factor measurements \cite{Laszlo.Zimboras:2018,Jentschura:2018,
  Ulbricht.Mueller.Surzhykov:2019,Ito:2021}, atom interferometry
\cite{Zych.EtAl:2011,Pikovski.EtAl:2015,Asenbaum.EtAl:2017,
  Loriani.EtAl:2019,Roura:2020,Roura.EtAl:2021,Lezeik.EtAl:2022}, and
metrology.

The relatively simple case of a single spin-half particle in an
external gravitational field that we dealt with here provides a good
example of the nature and degree on non-triviality immediately
encountered.  Given the many much further reaching claims that emerge
from various `approaches' to a theory of quantum gravity proper this
may be read as a call for some restraint.  On the other hand, just
listing longer and longer strings of corrections to Hamiltonians will
in the end also lead us nowhere without a consistent interpretational
scheme that eventually allows us to communicate the physical
significance of each term to our experimental colleagues.  In this
respect we tried hard to consistently stay within a well-defined
scheme, so as to produce each term of a given, well-defined order once
and only once.  In that respect we also wish to refer to our
discussion in \cite{Giulini.Grossardt.Schwartz:2022}.

Closest to our approach are the papers that we already discussed in
the introduction.  We claim to have improved on them concerning not
only the order of approximation but also concerning the systematics.
We showed that even within the larger (and hence more restricting) set
of approximation-hypotheses assumed in the most recent of these papers
\cite{Perche.Neuser:2021}, their list of terms for the final
Hamiltonian is not complete.  Ours, we believe, is.

Finally we wish to mention a characteristic difficulty concerning the
interpretation of interaction terms in Hamiltonians in the context of
General Relativity.  It has to do with the changing interpretation of
coordinates once the Hamiltonian refers to different metrics.  More
precisely, consider two Hamiltonian functions being given, one of
which takes into account the interaction with the gravitational field
to a higher degree than the other; then, strictly speaking, it is not
permissible to address the additional terms as the sole expression of
the higher order interaction, the reason being that together with the
higher degree of approximation to the metric, the metric meaning of
the coordinates, too, has also changed at the same time.  Again we
refer to \cite{Giulini.Grossardt.Schwartz:2022} for a more extensive
discussion, also providing a typical example.

\section*{Acknowledgements}

This work was supported by the Deutsche Forschungsgemeinschaft via the
Collaborative Research Centre 1227 `DQ-mat'---project number
274200144, project A05.

A.\,A.\ acknowledges funding from Trinity College, Cambridge via a
Rouse Ball Travelling Studentship.

\nocite{apsrev41Control}
\bibliography{Post-Newtonian_spinor,revtex-custom}

\begin{thebibliography}{37}%
\makeatletter
\providecommand \@ifxundefined [1]{%
 \@ifx{#1\undefined}
}%
\providecommand \@ifnum [1]{%
 \ifnum #1\expandafter \@firstoftwo
 \else \expandafter \@secondoftwo
 \fi
}%
\providecommand \@ifx [1]{%
 \ifx #1\expandafter \@firstoftwo
 \else \expandafter \@secondoftwo
 \fi
}%
\providecommand \natexlab [1]{#1}%
\providecommand \enquote  [1]{``#1''}%
\providecommand \bibnamefont  [1]{#1}%
\providecommand \bibfnamefont [1]{#1}%
\providecommand \citenamefont [1]{#1}%
\providecommand \href@noop [0]{\@secondoftwo}%
\providecommand \href [0]{\begingroup \@sanitize@url \@href}%
\providecommand \@href[1]{\@@startlink{#1}\@@href}%
\providecommand \@@href[1]{\endgroup#1\@@endlink}%
\providecommand \@sanitize@url [0]{\catcode `\\12\catcode `\$12\catcode
  `\&12\catcode `\#12\catcode `\^12\catcode `\_12\catcode `\%12\relax}%
\providecommand \@@startlink[1]{}%
\providecommand \@@endlink[0]{}%
\providecommand \url  [0]{\begingroup\@sanitize@url \@url }%
\providecommand \@url [1]{\endgroup\@href {#1}{\urlprefix }}%
\providecommand \urlprefix  [0]{URL }%
\providecommand \Eprint [0]{\href }%
\providecommand \doibase [0]{http://dx.doi.org/}%
\providecommand \selectlanguage [0]{\@gobble}%
\providecommand \bibinfo  [0]{\@secondoftwo}%
\providecommand \bibfield  [0]{\@secondoftwo}%
\providecommand \translation [1]{[#1]}%
\providecommand \BibitemOpen [0]{}%
\providecommand \bibitemStop [0]{}%
\providecommand \bibitemNoStop [0]{.\EOS\space}%
\providecommand \EOS [0]{\spacefactor3000\relax}%
\providecommand \BibitemShut  [1]{\csname bibitem#1\endcsname}%
\let\auto@bib@innerbib\@empty
\bibitem [{\citenamefont {L\'aszl\'o}\ and\ \citenamefont
  {Zimbor\'as}(2018)}]{Laszlo.Zimboras:2018}%
  \BibitemOpen
  \bibfield  {author} {\bibinfo {author} {\bibfnamefont {A.}~\bibnamefont
  {L\'aszl\'o}}\ and\ \bibinfo {author} {\bibfnamefont {Z.}~\bibnamefont
  {Zimbor\'as}},\ }\bibfield  {title} {\enquote {\bibinfo {title}
  {Quantification of {GR} effects in muon g-2, {EDM} and other spin precession
  experiments},}\ }\href {\doibase 10.1088/1361-6382/aacfee} {\bibfield
  {journal} {\bibinfo  {journal} {Class. Quantum Grav.}\ }\textbf {\bibinfo
  {volume} {35}},\ \bibinfo {pages} {175003} (\bibinfo {year} {2018})},\
  \Eprint {http://arxiv.org/abs/1803.01395}{1803.01395}\BibitemShut {NoStop}%
\bibitem [{\citenamefont {Jentschura}(2018)}]{Jentschura:2018}%
  \BibitemOpen
  \bibfield  {author} {\bibinfo {author} {\bibfnamefont {U.~D.}\ \bibnamefont
  {Jentschura}},\ }\bibfield  {title} {\enquote {\bibinfo {title}
  {Gravitational effects in $g$-factor measurements and high-preci\-sion
  spectroscopy: Limits of {Einstein's} equivalence principle},}\ }\href
  {\doibase 10.1103/PhysRevA.98.032508} {\bibfield  {journal} {\bibinfo
  {journal} {Phys. Rev. A}\ }\textbf {\bibinfo {volume} {98}},\ \bibinfo
  {pages} {032508} (\bibinfo {year} {2018})},\ \Eprint
  {http://arxiv.org/abs/1808.02089}{1808.02089}\BibitemShut {NoStop}%
\bibitem [{\citenamefont {Ulbricht}\ \emph {et~al.}(2019)\citenamefont
  {Ulbricht}, \citenamefont {M\"uller},\ and\ \citenamefont
  {Surzhykov}}]{Ulbricht.Mueller.Surzhykov:2019}%
  \BibitemOpen
  \bibfield  {author} {\bibinfo {author} {\bibfnamefont {S.}~\bibnamefont
  {Ulbricht}}, \bibinfo {author} {\bibfnamefont {R.~A.}\ \bibnamefont
  {M\"uller}}, \ and\ \bibinfo {author} {\bibfnamefont {A.}~\bibnamefont
  {Surzhykov}},\ }\bibfield  {title} {\enquote {\bibinfo {title} {Gravitational
  effects on geonium and free electron ${\mathrm{g}}_{s}$-factor measurements
  in a {Penning} trap},}\ }\href {\doibase 10.1103/PhysRevD.100.064029}
  {\bibfield  {journal} {\bibinfo  {journal} {Phys. Rev. D}\ }\textbf {\bibinfo
  {volume} {100}},\ \bibinfo {pages} {064029} (\bibinfo {year} {2019})},\
  \Eprint {http://arxiv.org/abs/1907.01460}{1907.01460}\BibitemShut {NoStop}%
\bibitem [{\citenamefont {Ito}(2021)}]{Ito:2021}%
  \BibitemOpen
  \bibfield  {author} {\bibinfo {author} {\bibfnamefont {A.}~\bibnamefont
  {Ito}},\ }\bibfield  {title} {\enquote {\bibinfo {title} {Inertial and
  gravitational effects on a geonium atom},}\ }\href {\doibase
  10.1088/1361-6382/ac1be9} {\bibfield  {journal} {\bibinfo  {journal} {Class.
  Quantum Grav.}\ }\textbf {\bibinfo {volume} {38}},\ \bibinfo {pages} {195015}
  (\bibinfo {year} {2021})},\ \Eprint
  {http://arxiv.org/abs/2011.11217}{2011.11217}\BibitemShut {NoStop}%
\bibitem [{\citenamefont {Micko}\ \emph {et~al.}(2022)\citenamefont {Micko},
  \citenamefont {Bosina}, \citenamefont {Cranganore}, \citenamefont {Jenke},
  \citenamefont {Pitschmann}, \citenamefont {Roccia}, \citenamefont {Sedmik},\
  and\ \citenamefont {Abele}}]{Micko.EtAl:2022}%
  \BibitemOpen
  \bibfield  {author} {\bibinfo {author} {\bibfnamefont {J.}~\bibnamefont
  {Micko}}, \bibinfo {author} {\bibfnamefont {J.}~\bibnamefont {Bosina}},
  \bibinfo {author} {\bibfnamefont {S.~S.}\ \bibnamefont {Cranganore}},
  \bibinfo {author} {\bibfnamefont {T.}~\bibnamefont {Jenke}}, \bibinfo
  {author} {\bibfnamefont {M.}~\bibnamefont {Pitschmann}}, \bibinfo {author}
  {\bibfnamefont {S.}~\bibnamefont {Roccia}}, \bibinfo {author} {\bibfnamefont
  {R.~I.~P.}\ \bibnamefont {Sedmik}}, \ and\ \bibinfo {author} {\bibfnamefont
  {H.}~\bibnamefont {Abele}},\ }\bibfield  {title} {\enquote {\bibinfo {title}
  {{qBounce: Systematic shifts of transition frequencies of gravitational
  states of ultra-cold neutrons using Ramsey gravity resonance
  spectroscopy}},}\ }in\ \href {\doibase 10.58027/1e1n-7973} {\emph {\bibinfo
  {booktitle} {Rencontres de Moriond 2022: Proceedings of the Gravitation
  Session, La Thuile, January 30-February 6 2022}}}\ (\bibinfo  {publisher}
  {ARISF},\ \bibinfo {year} {2022})\ pp.\ \bibinfo {pages} {143--148},\ \Eprint
  {http://arxiv.org/abs/2301.08583}{2301.08583}\BibitemShut {NoStop}%
\bibitem [{\citenamefont {Asenbaum}\ \emph {et~al.}(2017)\citenamefont
  {Asenbaum}, \citenamefont {Overstreet}, \citenamefont {Kovachy},
  \citenamefont {Brown}, \citenamefont {Hogan},\ and\ \citenamefont
  {Kasevich}}]{Asenbaum.EtAl:2017}%
  \BibitemOpen
  \bibfield  {author} {\bibinfo {author} {\bibfnamefont {P.}~\bibnamefont
  {Asenbaum}}, \bibinfo {author} {\bibfnamefont {C.}~\bibnamefont
  {Overstreet}}, \bibinfo {author} {\bibfnamefont {T.}~\bibnamefont {Kovachy}},
  \bibinfo {author} {\bibfnamefont {D.~D.}\ \bibnamefont {Brown}}, \bibinfo
  {author} {\bibfnamefont {J.~M.}\ \bibnamefont {Hogan}}, \ and\ \bibinfo
  {author} {\bibfnamefont {M.~A.}\ \bibnamefont {Kasevich}},\ }\bibfield
  {title} {\enquote {\bibinfo {title} {Phase shift in an atom interferometer
  due to spacetime curvature across its wave function},}\ }\href {\doibase
  10.1103/PhysRevLett.118.183602} {\bibfield  {journal} {\bibinfo  {journal}
  {Phys. Rev. Lett.}\ }\textbf {\bibinfo {volume} {118}},\ \bibinfo {pages}
  {183602} (\bibinfo {year} {2017})},\ \Eprint
  {http://arxiv.org/abs/1610.03832}{1610.03832}\BibitemShut {NoStop}%
\bibitem [{\citenamefont {Lezeik}\ \emph {et~al.}(2022)\citenamefont {Lezeik},
  \citenamefont {Tell}, \citenamefont {Zipfel}, \citenamefont {Gupta},
  \citenamefont {Wodey}, \citenamefont {Rasel}, \citenamefont {Schubert},\ and\
  \citenamefont {Schlippert}}]{Lezeik.EtAl:2022}%
  \BibitemOpen
  \bibfield  {author} {\bibinfo {author} {\bibfnamefont {A.}~\bibnamefont
  {Lezeik}}, \bibinfo {author} {\bibfnamefont {D.}~\bibnamefont {Tell}},
  \bibinfo {author} {\bibfnamefont {K.}~\bibnamefont {Zipfel}}, \bibinfo
  {author} {\bibfnamefont {V.}~\bibnamefont {Gupta}}, \bibinfo {author}
  {\bibfnamefont {{\'E}.}~\bibnamefont {Wodey}}, \bibinfo {author}
  {\bibfnamefont {E.~M.}\ \bibnamefont {Rasel}}, \bibinfo {author}
  {\bibfnamefont {C.}~\bibnamefont {Schubert}}, \ and\ \bibinfo {author}
  {\bibfnamefont {D.}~\bibnamefont {Schlippert}},\ }\href@noop {} {\enquote
  {\bibinfo {title} {Understanding the gravitational and magnetic environment
  of a very long baseline atom interferometer},}\ } (\bibinfo {year} {2022}),\
  \Eprint {http://arxiv.org/abs/2209.08886}{2209.08886}\BibitemShut {NoStop}%
\bibitem [{\citenamefont {Zych}\ \emph {et~al.}(2011)\citenamefont {Zych},
  \citenamefont {Costa}, \citenamefont {Pikovski},\ and\ \citenamefont
  {Brukner}}]{Zych.EtAl:2011}%
  \BibitemOpen
  \bibfield  {author} {\bibinfo {author} {\bibfnamefont {M.}~\bibnamefont
  {Zych}}, \bibinfo {author} {\bibfnamefont {F.}~\bibnamefont {Costa}},
  \bibinfo {author} {\bibfnamefont {I.}~\bibnamefont {Pikovski}}, \ and\
  \bibinfo {author} {\bibfnamefont {{\v C}.}~\bibnamefont {Brukner}},\
  }\bibfield  {title} {\enquote {\bibinfo {title} {Quantum interferometric
  visibility as a witness of general relativistic proper time},}\ }\href
  {\doibase 10.1038/ncomms1498} {\bibfield  {journal} {\bibinfo  {journal}
  {Nat. Commun.}\ }\textbf {\bibinfo {volume} {2}},\ \bibinfo {pages} {505}
  (\bibinfo {year} {2011})},\ \Eprint
  {http://arxiv.org/abs/1105.4531}{1105.4531}\BibitemShut {NoStop}%
\bibitem [{\citenamefont {Pikovski}\ \emph {et~al.}(2015)\citenamefont
  {Pikovski}, \citenamefont {Zych}, \citenamefont {Costa},\ and\ \citenamefont
  {Brukner}}]{Pikovski.EtAl:2015}%
  \BibitemOpen
  \bibfield  {author} {\bibinfo {author} {\bibfnamefont {I.}~\bibnamefont
  {Pikovski}}, \bibinfo {author} {\bibfnamefont {M.}~\bibnamefont {Zych}},
  \bibinfo {author} {\bibfnamefont {F.}~\bibnamefont {Costa}}, \ and\ \bibinfo
  {author} {\bibfnamefont {{\v C}.}~\bibnamefont {Brukner}},\ }\bibfield
  {title} {\enquote {\bibinfo {title} {Universal decoherence due to
  gravitational time dilation},}\ }\href {\doibase 10.1038/nphys3366}
  {\bibfield  {journal} {\bibinfo  {journal} {Nat. Phys.}\ }\textbf {\bibinfo
  {volume} {11}},\ \bibinfo {pages} {668--672} (\bibinfo {year} {2015})},\
  \Eprint {http://arxiv.org/abs/1311.1095}{1311.1095}\BibitemShut {NoStop}%
\bibitem [{\citenamefont {Schwartz}\ and\ \citenamefont
  {Giulini}(2019{\natexlab{a}})}]{Schwartz.Giulini:2019:AiG}%
  \BibitemOpen
  \bibfield  {author} {\bibinfo {author} {\bibfnamefont {P.~K.}\ \bibnamefont
  {Schwartz}}\ and\ \bibinfo {author} {\bibfnamefont {D.}~\bibnamefont
  {Giulini}},\ }\bibfield  {title} {\enquote {\bibinfo {title}
  {Post-{Newtonian} {Hamiltonian} description of an atom in a weak
  gravitational field},}\ }\href {\doibase 10.1103/PhysRevA.100.052116}
  {\bibfield  {journal} {\bibinfo  {journal} {Phys. Rev. A}\ }\textbf {\bibinfo
  {volume} {100}},\ \bibinfo {pages} {052116} (\bibinfo {year}
  {2019}{\natexlab{a}})},\ \Eprint
  {http://arxiv.org/abs/1908.06929}{1908.06929}\BibitemShut {NoStop}%
\bibitem [{\citenamefont {Schwartz}(2020)}]{Schwartz:2020}%
  \BibitemOpen
  \bibfield  {author} {\bibinfo {author} {\bibfnamefont {P.~K.}\ \bibnamefont
  {Schwartz}},\ }\emph {\bibinfo {title} {{Post-{Newtonian} Description of
  Quantum Systems in Gravitational Fields}}},\ \href {\doibase 10.15488/10085}
  {\bibinfo {type} {Doctoral thesis}},\ \bibinfo  {school} {Gottfried Wilhelm
  Leibniz Universit\"at Hannover} (\bibinfo {year} {2020}),\ \Eprint
  {http://arxiv.org/abs/2009.11319}{arXiv:2009.11319 [gr-qc]}\BibitemShut
  {NoStop}%
\bibitem [{\citenamefont {Loriani}\ \emph {et~al.}(2019)\citenamefont
  {Loriani}, \citenamefont {Friedrich}, \citenamefont {Ufrecht}, \citenamefont
  {Di~Pumpo}, \citenamefont {Kleinert}, \citenamefont {Abend}, \citenamefont
  {Gaaloul}, \citenamefont {Meiners}, \citenamefont {Schubert}, \citenamefont
  {Tell}, \citenamefont {Wodey}, \citenamefont {Zych}, \citenamefont {Ertmer},
  \citenamefont {Roura}, \citenamefont {Schlippert}, \citenamefont {Schleich},
  \citenamefont {Rasel},\ and\ \citenamefont {Giese}}]{Loriani.EtAl:2019}%
  \BibitemOpen
  \bibfield  {author} {\bibinfo {author} {\bibfnamefont {S.}~\bibnamefont
  {Loriani}}, \bibinfo {author} {\bibfnamefont {A.}~\bibnamefont {Friedrich}},
  \bibinfo {author} {\bibfnamefont {C.}~\bibnamefont {Ufrecht}}, \bibinfo
  {author} {\bibfnamefont {F.}~\bibnamefont {Di~Pumpo}}, \bibinfo {author}
  {\bibfnamefont {S.}~\bibnamefont {Kleinert}}, \bibinfo {author}
  {\bibfnamefont {S.}~\bibnamefont {Abend}}, \bibinfo {author} {\bibfnamefont
  {N.}~\bibnamefont {Gaaloul}}, \bibinfo {author} {\bibfnamefont
  {C.}~\bibnamefont {Meiners}}, \bibinfo {author} {\bibfnamefont
  {C.}~\bibnamefont {Schubert}}, \bibinfo {author} {\bibfnamefont
  {D.}~\bibnamefont {Tell}}, \bibinfo {author} {\bibfnamefont {E.}~\bibnamefont
  {Wodey}}, \bibinfo {author} {\bibfnamefont {M.}~\bibnamefont {Zych}},
  \bibinfo {author} {\bibfnamefont {W.}~\bibnamefont {Ertmer}}, \bibinfo
  {author} {\bibfnamefont {A.}~\bibnamefont {Roura}}, \bibinfo {author}
  {\bibfnamefont {D.}~\bibnamefont {Schlippert}}, \bibinfo {author}
  {\bibfnamefont {W.~P.}\ \bibnamefont {Schleich}}, \bibinfo {author}
  {\bibfnamefont {E.~M.}\ \bibnamefont {Rasel}}, \ and\ \bibinfo {author}
  {\bibfnamefont {E.}~\bibnamefont {Giese}},\ }\bibfield  {title} {\enquote
  {\bibinfo {title} {Interference of clocks: A quantum twin paradox},}\ }\href
  {\doibase 10.1126/sciadv.aax8966} {\bibfield  {journal} {\bibinfo  {journal}
  {Sci. Adv.}\ }\textbf {\bibinfo {volume} {5}},\ \bibinfo {pages} {eaax8966}
  (\bibinfo {year} {2019})},\ \Eprint
  {http://arxiv.org/abs/1905.09102}{1905.09102}\BibitemShut {NoStop}%
\bibitem [{\citenamefont {Roura}(2020)}]{Roura:2020}%
  \BibitemOpen
  \bibfield  {author} {\bibinfo {author} {\bibfnamefont {A.}~\bibnamefont
  {Roura}},\ }\bibfield  {title} {\enquote {\bibinfo {title} {Gravitational
  redshift in quantum-clock interferometry},}\ }\href {\doibase
  10.1103/PhysRevX.10.021014} {\bibfield  {journal} {\bibinfo  {journal} {Phys.
  Rev. X}\ }\textbf {\bibinfo {volume} {10}},\ \bibinfo {pages} {021014}
  (\bibinfo {year} {2020})},\ \Eprint
  {http://arxiv.org/abs/1810.06744}{1810.06744}\BibitemShut {NoStop}%
\bibitem [{\citenamefont {Roura}\ \emph {et~al.}(2021)\citenamefont {Roura},
  \citenamefont {Schubert}, \citenamefont {Schlippert},\ and\ \citenamefont
  {Rasel}}]{Roura.EtAl:2021}%
  \BibitemOpen
  \bibfield  {author} {\bibinfo {author} {\bibfnamefont {A.}~\bibnamefont
  {Roura}}, \bibinfo {author} {\bibfnamefont {C.}~\bibnamefont {Schubert}},
  \bibinfo {author} {\bibfnamefont {D.}~\bibnamefont {Schlippert}}, \ and\
  \bibinfo {author} {\bibfnamefont {E.~M.}\ \bibnamefont {Rasel}},\ }\bibfield
  {title} {\enquote {\bibinfo {title} {Measuring gravitational time dilation
  with delocalized quantum superpositions},}\ }\href {\doibase
  10.1103/PhysRevD.104.084001} {\bibfield  {journal} {\bibinfo  {journal}
  {Phys. Rev. D}\ }\textbf {\bibinfo {volume} {104}},\ \bibinfo {pages}
  {084001} (\bibinfo {year} {2021})},\ \Eprint
  {http://arxiv.org/abs/2010.11156}{2010.11156}\BibitemShut {NoStop}%
\bibitem [{\citenamefont {Manasse}\ and\ \citenamefont
  {Misner}(1963)}]{Manasse.Misner:1963}%
  \BibitemOpen
  \bibfield  {author} {\bibinfo {author} {\bibfnamefont {F.~K.}\ \bibnamefont
  {Manasse}}\ and\ \bibinfo {author} {\bibfnamefont {C.~W.}\ \bibnamefont
  {Misner}},\ }\bibfield  {title} {\enquote {\bibinfo {title} {Fermi normal
  coordinates and some basic concepts in differential geometry},}\ }\href
  {\doibase 10.1063/1.1724316} {\bibfield  {journal} {\bibinfo  {journal} {J.
  Math. Phys.}\ }\textbf {\bibinfo {volume} {4}},\ \bibinfo {pages} {735--745}
  (\bibinfo {year} {1963})}\BibitemShut {NoStop}%
\bibitem [{\citenamefont {Ni}\ and\ \citenamefont
  {Zimmermann}(1978)}]{Ni.Zimmermann:1978}%
  \BibitemOpen
  \bibfield  {author} {\bibinfo {author} {\bibfnamefont {W.-T.}\ \bibnamefont
  {Ni}}\ and\ \bibinfo {author} {\bibfnamefont {M.}~\bibnamefont
  {Zimmermann}},\ }\bibfield  {title} {\enquote {\bibinfo {title} {Inertial and
  gravitational effects in the proper reference frame of an accelerated,
  rotating observer},}\ }\href {\doibase 10.1103/PhysRevD.17.1473} {\bibfield
  {journal} {\bibinfo  {journal} {Phys. Rev. D}\ }\textbf {\bibinfo {volume}
  {17}},\ \bibinfo {pages} {1473--1476} (\bibinfo {year} {1978})}\BibitemShut
  {NoStop}%
\bibitem [{\citenamefont {Li}\ and\ \citenamefont {Ni}(1979)}]{Li.Ni:1979}%
  \BibitemOpen
  \bibfield  {author} {\bibinfo {author} {\bibfnamefont {W.-Q.}\ \bibnamefont
  {Li}}\ and\ \bibinfo {author} {\bibfnamefont {W.-T.}\ \bibnamefont {Ni}},\
  }\bibfield  {title} {\enquote {\bibinfo {title} {Coupled inertial and
  gravitational effects in the proper reference frame of an accelerated,
  rotating observer},}\ }\href {\doibase 10.1063/1.524203} {\bibfield
  {journal} {\bibinfo  {journal} {J. Math. Phys.}\ }\textbf {\bibinfo {volume}
  {20}},\ \bibinfo {pages} {1473--1480} (\bibinfo {year} {1979})}\BibitemShut
  {NoStop}%
\bibitem [{\citenamefont {Giulini}\ and\ \citenamefont
  {Gro{\ss}ardt}(2012)}]{Giulini.Grossardt:2012}%
  \BibitemOpen
  \bibfield  {author} {\bibinfo {author} {\bibfnamefont {D.}~\bibnamefont
  {Giulini}}\ and\ \bibinfo {author} {\bibfnamefont {A.}~\bibnamefont
  {Gro{\ss}ardt}},\ }\bibfield  {title} {\enquote {\bibinfo {title} {The
  {Schr\"odinger--Newton} equation as a non-relativistic limit of
  self-gravitating {Klein--Gordon} and {Dirac} fields},}\ }\href {\doibase
  10.1088/0264-9381/29/21/215010} {\bibfield  {journal} {\bibinfo  {journal}
  {Class. Quantum Grav.}\ }\textbf {\bibinfo {volume} {29}},\ \bibinfo {pages}
  {215010} (\bibinfo {year} {2012})},\ \Eprint
  {http://arxiv.org/abs/1206.4250}{1206.4250}\BibitemShut {NoStop}%
\bibitem [{\citenamefont {Schwartz}\ and\ \citenamefont
  {Giulini}(2019{\natexlab{b}})}]{Schwartz.Giulini:2019:PNSE}%
  \BibitemOpen
  \bibfield  {author} {\bibinfo {author} {\bibfnamefont {P.~K.}\ \bibnamefont
  {Schwartz}}\ and\ \bibinfo {author} {\bibfnamefont {D.}~\bibnamefont
  {Giulini}},\ }\bibfield  {title} {\enquote {\bibinfo {title}
  {Post-{Newtonian} corrections to {Schr\"o\-din\-ger} equations in
  gravitational fields},}\ }\href {\doibase 10.1088/1361-6382/ab0fbd}
  {\bibfield  {journal} {\bibinfo  {journal} {Class. Quantum Grav.}\ }\textbf
  {\bibinfo {volume} {36}},\ \bibinfo {pages} {095016} (\bibinfo {year}
  {2019}{\natexlab{b}})},\ \bibinfo {note} {corrigendum published in
  \href{https://doi.org/10.1088/1361-6382/ab5633}{Class. Quantum Grav.
  \textbf{36} (2019), 249502}},\ \Eprint
  {http://arxiv.org/abs/1812.05181}{1812.05181}\BibitemShut {NoStop}%
\bibitem [{\citenamefont {Giulini}\ \emph {et~al.}(2023)\citenamefont
  {Giulini}, \citenamefont {Gro{\ss}ardt},\ and\ \citenamefont
  {Schwartz}}]{Giulini.Grossardt.Schwartz:2022}%
  \BibitemOpen
  \bibfield  {author} {\bibinfo {author} {\bibfnamefont {D.}~\bibnamefont
  {Giulini}}, \bibinfo {author} {\bibfnamefont {A.}~\bibnamefont
  {Gro{\ss}ardt}}, \ and\ \bibinfo {author} {\bibfnamefont {P.~K.}\
  \bibnamefont {Schwartz}},\ }\enquote {\bibinfo {title} {Coupling quantum
  matter and gravity},}\ in\ \href {\doibase 10.1007/978-3-031-31520-6_16}
  {\emph {\bibinfo {booktitle} {Modified and Quantum Gravity}}},\ \bibinfo
  {series} {Lecture Notes in Physics}, Vol.\ \bibinfo {volume} {1017},\
  \bibinfo {editor} {edited by\ \bibinfo {editor} {\bibfnamefont
  {C.}~\bibnamefont {Pfeifer}}\ and\ \bibinfo {editor} {\bibfnamefont
  {C.}~\bibnamefont {Lämmerzahl}}}\ (\bibinfo  {publisher} {Springer},\
  \bibinfo {address} {Cham},\ \bibinfo {year} {2023})\ Chap.~\bibinfo {chapter}
  {16},\ \Eprint {http://arxiv.org/abs/2207.05029}{2207.05029}\BibitemShut
  {NoStop}%
\bibitem [{\citenamefont {Parker}(1980{\natexlab{a}})}]{Parker:1980:PRL}%
  \BibitemOpen
  \bibfield  {author} {\bibinfo {author} {\bibfnamefont {L.}~\bibnamefont
  {Parker}},\ }\bibfield  {title} {\enquote {\bibinfo {title} {One-electron
  atom in curved space-time},}\ }\href {\doibase 10.1103/PhysRevLett.44.1559}
  {\bibfield  {journal} {\bibinfo  {journal} {Phys. Rev. Lett.}\ }\textbf
  {\bibinfo {volume} {44}},\ \bibinfo {pages} {1559--1562} (\bibinfo {year}
  {1980}{\natexlab{a}})}\BibitemShut {NoStop}%
\bibitem [{\citenamefont {Parker}(1980{\natexlab{b}})}]{Parker:1980:PRD}%
  \BibitemOpen
  \bibfield  {author} {\bibinfo {author} {\bibfnamefont {L.}~\bibnamefont
  {Parker}},\ }\bibfield  {title} {\enquote {\bibinfo {title} {One-electron
  atom as a probe of spacetime curvature},}\ }\href {\doibase
  10.1103/PhysRevD.22.1922} {\bibfield  {journal} {\bibinfo  {journal} {Phys.
  Rev. D}\ }\textbf {\bibinfo {volume} {22}},\ \bibinfo {pages} {1922--1934}
  (\bibinfo {year} {1980}{\natexlab{b}})}\BibitemShut {NoStop}%
\bibitem [{\citenamefont {Foldy}\ and\ \citenamefont
  {Wouthuysen}(1950)}]{Foldy.Wouthuysen:1950}%
  \BibitemOpen
  \bibfield  {author} {\bibinfo {author} {\bibfnamefont {L.~L.}\ \bibnamefont
  {Foldy}}\ and\ \bibinfo {author} {\bibfnamefont {S.~A.}\ \bibnamefont
  {Wouthuysen}},\ }\bibfield  {title} {\enquote {\bibinfo {title} {On the
  {Dirac} theory of spin 1/2 particles and its non-relativistic limit},}\
  }\href {\doibase 10.1103/PhysRev.78.29} {\bibfield  {journal} {\bibinfo
  {journal} {Phys. Rev.}\ }\textbf {\bibinfo {volume} {78}},\ \bibinfo {pages}
  {29--36} (\bibinfo {year} {1950})}\BibitemShut {NoStop}%
\bibitem [{\citenamefont {Perche}\ and\ \citenamefont
  {Neuser}(2021)}]{Perche.Neuser:2021}%
  \BibitemOpen
  \bibfield  {author} {\bibinfo {author} {\bibfnamefont {T.~R.}\ \bibnamefont
  {Perche}}\ and\ \bibinfo {author} {\bibfnamefont {J.}~\bibnamefont
  {Neuser}},\ }\bibfield  {title} {\enquote {\bibinfo {title} {A wavefunction
  description for a localized quantum particle in curved spacetimes},}\ }\href
  {\doibase 10.1088/1361-6382/ac103d} {\bibfield  {journal} {\bibinfo
  {journal} {Class. Quantum Gravity}\ }\textbf {\bibinfo {volume} {38}},\
  \bibinfo {pages} {175002} (\bibinfo {year} {2021})},\ \Eprint
  {http://arxiv.org/abs/2012.08539}{2012.08539}\BibitemShut {NoStop}%
\bibitem [{\citenamefont {Alibabaei}(2022)}]{Alibabaei:2022Thesis}%
  \BibitemOpen
  \bibfield  {author} {\bibinfo {author} {\bibfnamefont {A.}~\bibnamefont
  {Alibabaei}},\ }\emph {\bibinfo {title} {Geometric post-{Newtonian}
  description of spin-half particles in curved spacetime}},\ \href@noop {}
  {Master's thesis},\ \bibinfo  {school} {Gottfried Wilhelm Leibniz
  Universit\"at Hannover} (\bibinfo {year} {2022}),\ \Eprint
  {http://arxiv.org/abs/2204.05997}{2204.05997}\BibitemShut {NoStop}%
\bibitem [{\citenamefont {Geroch}(1968)}]{Geroch:1968}%
  \BibitemOpen
  \bibfield  {author} {\bibinfo {author} {\bibfnamefont {R.}~\bibnamefont
  {Geroch}},\ }\bibfield  {title} {\enquote {\bibinfo {title} {Spinor structure
  of space‐times in general relativity. {I}},}\ }\href {\doibase
  10.1063/1.1664507} {\bibfield  {journal} {\bibinfo  {journal} {J. Math.
  Phys.}\ }\textbf {\bibinfo {volume} {9}},\ \bibinfo {pages} {1739--1744}
  (\bibinfo {year} {1968})}\BibitemShut {NoStop}%
\bibitem [{\citenamefont {Collas}\ and\ \citenamefont
  {Klein}(2019)}]{Collas.Klein:2019}%
  \BibitemOpen
  \bibfield  {author} {\bibinfo {author} {\bibfnamefont {P.}~\bibnamefont
  {Collas}}\ and\ \bibinfo {author} {\bibfnamefont {D.}~\bibnamefont {Klein}},\
  }\href {\doibase 10.1007/978-3-030-14825-6} {\emph {\bibinfo {title} {{The
  Dirac Equation in Curved Spacetime}}}},\ SpringerBriefs in Physics\ (\bibinfo
   {publisher} {Springer},\ \bibinfo {address} {Cham},\ \bibinfo {year}
  {2019})\ \Eprint {http://arxiv.org/abs/1809.02764}{1809.02764}\BibitemShut
  {NoStop}%
\bibitem [{\citenamefont {Inonu}\ and\ \citenamefont
  {Wigner}(1953)}]{Inonu.Wigner:1953}%
  \BibitemOpen
  \bibfield  {author} {\bibinfo {author} {\bibfnamefont {E.}~\bibnamefont
  {Inonu}}\ and\ \bibinfo {author} {\bibfnamefont {E.~P.}\ \bibnamefont
  {Wigner}},\ }\bibfield  {title} {\enquote {\bibinfo {title} {On the
  contraction of groups and their representations},}\ }\href {\doibase
  10.1073/pnas.39.6.510} {\bibfield  {journal} {\bibinfo  {journal} {Proc.
  Natl. Acad. Sci. U.S.A.}\ }\textbf {\bibinfo {volume} {39}},\ \bibinfo
  {pages} {510--524} (\bibinfo {year} {1953})}\BibitemShut {NoStop}%
\bibitem [{\citenamefont {Dautcourt}(1997)}]{Dautcourt:1997}%
  \BibitemOpen
  \bibfield  {author} {\bibinfo {author} {\bibfnamefont {G.}~\bibnamefont
  {Dautcourt}},\ }\bibfield  {title} {\enquote {\bibinfo {title}
  {Post-{Newtonian} extension of the {Newton--Cartan} theory},}\ }\href
  {\doibase 10.1088/0264-9381/14/1A/009} {\bibfield  {journal} {\bibinfo
  {journal} {Class. Quantum Gravity}\ }\textbf {\bibinfo {volume} {14}},\
  \bibinfo {pages} {A109} (\bibinfo {year} {1997})},\ \Eprint
  {http://arxiv.org/abs/gr-qc/9610036}{gr-qc/9610036}\BibitemShut {NoStop}%
\bibitem [{\citenamefont {Tichy}\ and\ \citenamefont
  {Flanagan}(2011)}]{Tichy.Flanagan:2011}%
  \BibitemOpen
  \bibfield  {author} {\bibinfo {author} {\bibfnamefont {W.}~\bibnamefont
  {Tichy}}\ and\ \bibinfo {author} {\bibfnamefont {{\'E}.~{\'E}.}\ \bibnamefont
  {Flanagan}},\ }\bibfield  {title} {\enquote {\bibinfo {title} {Covariant
  formulation of the post-1-{Newtonian} approximation to general relativity},}\
  }\href {\doibase 10.1103/PhysRevD.84.044038} {\bibfield  {journal} {\bibinfo
  {journal} {Phys. Rev. D}\ }\textbf {\bibinfo {volume} {84}},\ \bibinfo
  {pages} {044038} (\bibinfo {year} {2011})},\ \Eprint
  {http://arxiv.org/abs/1101.0588}{1101.0588}\BibitemShut {NoStop}%
\bibitem [{\citenamefont {Hansen}\ \emph {et~al.}(2019)\citenamefont {Hansen},
  \citenamefont {Hartong},\ and\ \citenamefont {Obers}}]{Hansen.EtAl:2019}%
  \BibitemOpen
  \bibfield  {author} {\bibinfo {author} {\bibfnamefont {D.}~\bibnamefont
  {Hansen}}, \bibinfo {author} {\bibfnamefont {J.}~\bibnamefont {Hartong}}, \
  and\ \bibinfo {author} {\bibfnamefont {N.~A.}\ \bibnamefont {Obers}},\
  }\bibfield  {title} {\enquote {\bibinfo {title} {Action principle for
  {Newtonian} gravity},}\ }\href {\doibase 10.1103/PhysRevLett.122.061106}
  {\bibfield  {journal} {\bibinfo  {journal} {Phys. Rev. Lett.}\ }\textbf
  {\bibinfo {volume} {122}},\ \bibinfo {pages} {061106} (\bibinfo {year}
  {2019})},\ \Eprint {http://arxiv.org/abs/1807.04765}{1807.04765}\BibitemShut
  {NoStop}%
\bibitem [{\citenamefont {Hansen}\ \emph {et~al.}(2020)\citenamefont {Hansen},
  \citenamefont {Hartong},\ and\ \citenamefont {Obers}}]{Hansen.EtAl:2020}%
  \BibitemOpen
  \bibfield  {author} {\bibinfo {author} {\bibfnamefont {D.}~\bibnamefont
  {Hansen}}, \bibinfo {author} {\bibfnamefont {J.}~\bibnamefont {Hartong}}, \
  and\ \bibinfo {author} {\bibfnamefont {N.~A.}\ \bibnamefont {Obers}},\
  }\bibfield  {title} {\enquote {\bibinfo {title} {Non-relativistic gravity and
  its coupling to matter},}\ }\href {\doibase 10.1007/JHEP06(2020)145}
  {\bibfield  {journal} {\bibinfo  {journal} {J. High Energy Phys.}\ }\textbf
  {\bibinfo {volume} {2020}},\ \bibinfo {pages} {145} (\bibinfo {year}
  {2020})},\ \Eprint {http://arxiv.org/abs/2001.10277}{2001.10277}\BibitemShut
  {NoStop}%
\bibitem [{\citenamefont {Hartong}\ \emph {et~al.}(2023)\citenamefont
  {Hartong}, \citenamefont {Obers},\ and\ \citenamefont
  {Oling}}]{Hartong.EtAl:2023}%
  \BibitemOpen
  \bibfield  {author} {\bibinfo {author} {\bibfnamefont {J.}~\bibnamefont
  {Hartong}}, \bibinfo {author} {\bibfnamefont {N.~A.}\ \bibnamefont {Obers}},
  \ and\ \bibinfo {author} {\bibfnamefont {G.}~\bibnamefont {Oling}},\
  }\bibfield  {title} {\enquote {\bibinfo {title} {Review on non-relativistic
  gravity},}\ }\href {\doibase 10.3389/fphy.2023.1116888} {\bibfield  {journal}
  {\bibinfo  {journal} {Front. Phys.}\ }\textbf {\bibinfo {volume} {11}},\
  \bibinfo {pages} {1116888} (\bibinfo {year} {2023})},\ \Eprint
  {http://arxiv.org/abs/2212.11309}{2212.11309}\BibitemShut {NoStop}%
\bibitem [{\citenamefont {Schwartz}(2023)}]{Schwartz:2023}%
  \BibitemOpen
  \bibfield  {author} {\bibinfo {author} {\bibfnamefont {P.~K.}\ \bibnamefont
  {Schwartz}},\ }\bibfield  {title} {\enquote {\bibinfo {title} {Teleparallel
  {Newton--Cartan} gravity},}\ }\href {\doibase 10.1088/1361-6382/accc02}
  {\bibfield  {journal} {\bibinfo  {journal} {Class. Quantum Gravity}\ }\textbf
  {\bibinfo {volume} {40}},\ \bibinfo {pages} {105008} (\bibinfo {year}
  {2023})},\ \Eprint {http://arxiv.org/abs/2211.11796}{2211.11796}\BibitemShut
  {NoStop}%
\bibitem [{\citenamefont {Wald}(1994)}]{Wald:1994}%
  \BibitemOpen
  \bibfield  {author} {\bibinfo {author} {\bibfnamefont {R.~M.}\ \bibnamefont
  {Wald}},\ }\href@noop {} {\emph {\bibinfo {title} {{Quantum Field Theory in
  Curved Spacetime and Black Hole Thermodynamics}}}},\ Chicago Lectures in
  Physics\ (\bibinfo  {publisher} {University of Chicago Press},\ \bibinfo
  {address} {Chicago},\ \bibinfo {year} {1994})\BibitemShut {NoStop}%
\bibitem [{\citenamefont {Newton}\ and\ \citenamefont
  {Wigner}(1949)}]{Newton.Wigner:1949}%
  \BibitemOpen
  \bibfield  {author} {\bibinfo {author} {\bibfnamefont {T.~D.}\ \bibnamefont
  {Newton}}\ and\ \bibinfo {author} {\bibfnamefont {E.~P.}\ \bibnamefont
  {Wigner}},\ }\bibfield  {title} {\enquote {\bibinfo {title} {Localized states
  for elementary systems},}\ }\href {\doibase 10.1103/revmodphys.21.400}
  {\bibfield  {journal} {\bibinfo  {journal} {Rev. Mod. Phys.}\ }\textbf
  {\bibinfo {volume} {21}},\ \bibinfo {pages} {400--406} (\bibinfo {year}
  {1949})}\BibitemShut {NoStop}%
\bibitem [{\citenamefont {Schwartz}\ and\ \citenamefont
  {Giulini}(2020)}]{Schwartz.Giulini:2020}%
  \BibitemOpen
  \bibfield  {author} {\bibinfo {author} {\bibfnamefont {P.~K.}\ \bibnamefont
  {Schwartz}}\ and\ \bibinfo {author} {\bibfnamefont {D.}~\bibnamefont
  {Giulini}},\ }\bibfield  {title} {\enquote {\bibinfo {title} {Classical
  perspectives on the {Newton--Wigner} position observable},}\ }\href {\doibase
  10.1142/S0219887820501765} {\bibfield  {journal} {\bibinfo  {journal} {Int.
  J. Geom. Methods Mod. Phys.}\ }\textbf {\bibinfo {volume} {17}},\ \bibinfo
  {pages} {2050176} (\bibinfo {year} {2020})},\ \Eprint
  {http://arxiv.org/abs/2004.09723}{2004.09723}\BibitemShut {NoStop}%
\end{thebibliography}%

\appendix

\section{The connection in generalised Fermi normal coordinates}
\label{sec:app_conn_FN}

The Christoffel symbols in generalised Fermi normal coordinates were
calculated to second order in the geodesic distance to the reference
worldline in \cite{Li.Ni:1979}.  Note that in this reference, some
calculational errors were made, which we have corrected in the
following and marked in red.  The Christoffel symbols are given by
\begin{subequations} \label{eq:Christoffel_FNC}
\begin{align}
  \Gamma^s_{ss}
  &= c^{-3} (\vect b \cdot \vect x
      + 2 \vect a \cdot (\vect\omega \times \vect x))
    + \frac{1}{2} R_{0l0m;0} x^l x^m
    + \frac{1}{3} c^{-2} a^i R_{0lim} x^l x^m \nonumber\\
  &\quad - c^{-5} (\vect b \cdot \vect x
        + 2 \vect a \cdot (\vect\omega \times \vect x))
      (\vect a \cdot \vect x)
    \textcolor{red}{{}+ 2 c^{-1} R_{0i0j}
      (\vect\omega \times \vect x)^i x^j}
    + \Or(x^3),\\
  \Gamma^s_{si}
  &= c^{-2} a_i - c^{-4} a_i (\vect a \cdot \vect x)
    + R_{0i0j} x^j
    \mathbin{\textcolor{red}{+}} \frac{1}{6}
      (R_{0l0m;i} + 2 R_{0i0l;m}) x^l x^m
    - \frac{2}{3} c^{-2} (\vect a \cdot \vect x) R_{0i0j} x^j
    \nonumber\\
  &\quad - \frac{1}{3} c^{-2} a_i R_{0l0m} x^l x^m
    + c^{-6} a_i (\vect a \cdot \vect x)^2
    - \frac{1}{3} c^{-1} (\vect\omega \times \vect x)^k
      (R_{0ilk} + R_{0kli}) x^l + \Or(x^3),\\
  \Gamma^s_{ij}
  &= \frac{1}{3} \left\{2 R_{0(ij)k}
    + \frac{1}{4} (5 R_{0(ij)k;l} - R_{0kl(i;j)}) x^l
    - 2 c^{-2} (\vect a \cdot \vect x) R_{0(ij)k}
    \right\} x^k + \Or(x^3),\\
  \Gamma^i_{ss}
  &= c^{-2} a^i + \tensor{R}{_0^i_{0j}} x^j
    + c^{-2} (\vect\eta \times \vect x)^i
    + c^{-4} (\vect a \cdot \vect x) a^i
    + c^{-2} (\vect\omega \times (\vect\omega \times \vect x))^i
    - \frac{1}{2} \tensor{R}{_{0l0m}^{;i}} x^l x^m
    \nonumber\\
  &\quad + \tensor{R}{_0^i_{0l;m}} x^l x^m
    + 2 c^{-2} (\vect a \cdot \vect x) \tensor{R}{_0^i_{0j}} x^j
    - \frac{1}{3} c^{-2} a^j \tensor{R}{^i_{ljm}} x^l x^m \nonumber\\
  &\quad - c^{-4} (\vect\omega \times \vect x)^i (\vect b \cdot \vect x
      + 2 \vect a \cdot (\vect\omega \times \vect x))
    - 2 c^{-1} (\vect\omega \times \vect x)^k
      \tensor{R}{_{0j}^i_k} x^j
    + \Or(x^3), \displaybreak[0]\\
  \Gamma^i_{sj}
  &= - c^{-1} \tensor{\varepsilon}{^i_{jk}} \omega^k
    - \tensor{R}{_{0k}^i_j} x^k
    - c^{-3} (\vect\omega \times \vect x)^i a_j
    + \left\{ \mathbin{\textcolor{red}{+}} \frac{1}{6}
        \tensor{R}{_{0j}^i_{l;m}}
      - \frac{1}{2} \tensor{R}{_{0l}^i_{j;m}}
      \mathbin{\textcolor{red}{-}} \frac{1}{6}
        \tensor{R}{_{0l}^i_{m;j}}
      \right\} x^l x^m \nonumber\\
  &\quad - \frac{1}{3} c^{-2} (\vect a \cdot \vect x)
      (\tensor{R}{_{0k}^i_j} + \tensor{R}{_0^i_{kj}}) x^k
    + \frac{1}{3} c^{-2} a_j \tensor{R}{_{0l}^i_m} x^l x^m
    - c^{-1} (\vect\omega \times \vect x)^i R_{0j0k} x^k \nonumber\\
  &\quad - \frac{1}{3} c^{-1} (\vect\omega \times \vect x)^l
      (\tensor{R}{_{lk}^i_j} + \tensor{R}{_l^i_{kj}}) x^k
    + c^{-5} a_j (\vect a \cdot \vect x)(\vect\omega \times \vect x)^i
    + \Or(x^3),\\
  \Gamma^i_{jk}
  &= - \frac{1}{3} \left\{2 \tensor{R}{^i_{(jk)l}}
    + \frac{1}{4} (5 \tensor{R}{^i_{(jk)l;m}}
      - \tensor{R}{^i_{lm(j;k)}}) x^m
    + 2 c^{-1} (\vect\omega \times \vect x)^i R_{0(jk)l} \right\} x^l
    + \Or(x^3).
\end{align}
\end{subequations}
The local connection form with respect to the frame \eqref{eq:frame}
is given by
\begin{subequations} \label{eq:local_conn_form_FNC}
\begin{align}
  \tensor{\omega}{_\mu^0_0}
  &= 0,\\
  \tensor{\omega}{_s^0_i}
  &= c^{-2} a_i + R_{0i0l} x^l
    + \frac{1}{2} c^{-2} (\vect a \cdot \vect x) R_{0i0l} x^l
    + \frac{1}{2} c^{-1} (\vect\omega \times \vect x)^k R_{0ikl} x^l
    \nonumber\\
  &\quad + \frac{1}{2} R_{0i0l;m} x^l x^m
    + \Or(x^3),\\
  \tensor{\omega}{_i^0_j}
  &= \frac{1}{2} R_{0jil} x^l
    + \frac{1}{3} R_{0jil;m} x^l x^m
    + \Or(x^3),\\
  \tensor{\omega}{_\mu^i_0}
  &= \delta^{ij} \tensor{\omega}{_\mu^0_j} \; ,\\
  \tensor{\omega}{_s^i_j}
  &= - c^{-1} \tensor{\varepsilon}{^i_{jk}} \omega^k
    - \tensor{R}{^i_{j0l}} x^l
    - \frac{1}{2} c^{-2} (\vect a \cdot \vect x) \tensor{R}{^i_{j0l}}
      x^l
    - \frac{1}{2} c^{-1} (\vect\omega \times \vect x)^k
      \tensor{R}{^i_{jkl}} x^l \nonumber\\
  &\quad - \frac{1}{2} \tensor{R}{^i_{j0l;m}} x^l x^m
    + \Or(x^3),\\
  \tensor{\omega}{_k^i_j}
  &= - \frac{1}{2} \tensor{R}{^i_{jkl}} x^l
    - \frac{1}{3} \tensor{R}{^i_{jkl;m}} x^l x^m
    + \Or(x^3).
\end{align}
\end{subequations}

\section{Stationarity with respect to the generalised Fermi normal
  coordinate time translation field}
\label{sec:app_stat_geom}

In the following, we are going to briefly discuss the geometric
interpretation of the possible condition that the metric be stationary
with respect to the time coordinate $\tau$ of the generalised Fermi
normal coordinates introduced in \cref{sec:expansion_FNC}, i.e.\ that
the timelike vector field $\partial/\partial\tau$ be Killing.  Note
that, as explained in the main text, the post-Newtonian expansion in
$c^{-1}$ of \cref{sec:expansion_post-Newtonian} is \emph{still} a
meaningful approximation procedure if stationarity does \emph{not}
hold, formulating the Dirac theory as a deformation of its Newtonian
limit.

As a first step, stationarity with respect to $\partial/\partial\tau$
of course means that the reference worldline $\gamma$ has to be
stationary.

Away from the worldline, $\partial/\partial\tau$ being Killing means
that the metric components \eqref{eq:metric_FNC} need be independent
of coordinate time $\tau$; i.e.\ we need the components $a^i$,
$\omega^i$, $R_{IJKL}$ of the acceleration of $\gamma$, the angular
velocity of the spatial basis $(\e_i)$ and the curvature to be
constant along the reference worldline $\gamma$:
\begin{equation}
  \dot a^i(\tau) = 0, \quad
  \dot\omega^i(\tau) = 0, \quad
  \dot R_{IJKL}(\tau) = 0
\end{equation}
Note, however, that the components are taken \emph{with respect to the
  generalised Fermi normal coordinates}; therefore, to see the true
geometric meaning of these conditions, we need to rewrite them
covariantly.

By direct computation, for the covariant derivatives of acceleration
and angular velocity we have
\begin{align}
  b^i(\tau)
  &= (\nabla_{\dot\gamma(\tau)} a(\tau))^i
    = \dot a^i(\tau) + c \Gamma^i_{sj}(\gamma(\tau)) a^j(\tau)
    = \dot a^i(\tau) + (\vect\omega(\tau) \times \vect a(\tau))^i ,
  \\
  \eta^i(\tau)
  &= (\nabla_{\dot\gamma(\tau)} \omega(\tau))^i
    = \dot \omega^i(\tau) + c \Gamma^i_{sj}(\gamma(\tau))
      \omega^j(\tau)
    = \dot \omega^i(\tau).
\end{align}
Thus, we see that stationarity of the metric with respect to the time
translation vector field given by generalised Fermi normal coordinates
implies that the angular velocity $\omega$ of the spatial reference
vectors be covariantly constant along the reference worldline
$\gamma$.  However, in the case of $\omega$ being non-zero, in the
generic case the worldline's acceleration $a$ need \emph{not} be
covariantly constant---it has to itself rotate with angular velocity
$\omega$, such that its components with respect to the rotating basis
are constant.  This may sound somewhat artificial, but note that for
example one could satisfy this condition with a covariantly constant
acceleration $a$ and spatial basis vectors $(\e_i)$ that rotate around
the axis given by $a$.

Of course, the condition of constancy of the curvature components
along $\gamma$ can be rewritten in terms of covariant derivatives of
the curvature tensor as well; however, this does not lead to any great
insight, so we will refrain from doing so here.

\section{The Dirac Hamiltonian up to \texorpdfstring{$\Or(c^{-1}x^4) +
    \Or(c^{-2}x^3)$}{O(c⁻¹x⁴) + O(c⁻²x³)}}
\label{sec:app_Dirac_higher_order}

In the main text, for a consistent post-Newtonian expansion of the
Dirac Hamiltonian leading to a resulting Pauli Hamiltonian known to
order $c^{-2}$ and $x^2$, we need to know the Dirac Hamiltonian to
order $x^3$ in those terms of order up to $c^{-1}$ in the
$c^{-1}$-expansion.  Going through the derivation of
\cite{Li.Ni:1979}, one can convince oneself that all $x$-dependent
terms in the Christoffel symbols in generalised Fermi normal
coordinates are of order at least $c^{-2}$ when expanding also in
$c^{-1}$; and employing the methods from \cite{Li.Ni:1979}, one can go
to higher order and calculate the order-$x^3$ terms to order $c^{-2}$.
The resulting Christoffel symbols read as follows, with the newly
calculated terms marked in blue (note that we use the ordering of
terms and the $\Or(c^{-n} x^m)$ notation as explained in the main text
before \eqref{eq:explain_sorting_double_exp}):
\begin{subequations} \label{eq:Christoffel_FNC_higher_order}
\begin{align}
  \Gamma^s_{ss}
  &= c^{-2} \left(\frac{c^2}{2} R_{0l0m;0} x^l x^m
      \textcolor{blue}{{}+ \frac{c^2}{6} R_{0l0m;n0} x^l x^m x^n}
    \right)
    \nonumber\\
  &\quad+ c^{-3} (\vect b \cdot \vect x
      + 2 \vect a \cdot (\vect\omega \times \vect x)
      + 2 c^2 R_{0i0j} (\vect\omega \times \vect x)^i x^j)
    + c^{-4} \left(\frac{c^2}{3} a^i R_{0lim} x^l x^m \right)
    \nonumber\\
  &\quad- c^{-5} \left( (\vect b \cdot \vect x
        + 2 \vect a \cdot (\vect\omega \times \vect x))
      (\vect a \cdot \vect x)\right)
    + \Or(c^{-2}x^4) + \Or(c^{-3}x^3), \displaybreak[0]\\
  \Gamma^s_{si}
  &= c^{-2} \left(a_i + c^2 R_{0i0j} x^j
      + \frac{c^2}{6} (R_{0l0m;i} + 2 R_{0i0l;m}) x^l x^m
      \textcolor{blue}{{}+ \frac{c^2}{12} (R_{0i0l;mn} + R_{0l0m;ni})
        x^l x^m x^n}\right) \nonumber\\
  &\quad+ c^{-3} \left(- \frac{c^2}{3} (\vect\omega \times \vect x)^k
      (R_{0ilk} + R_{0kli}) x^l \right)
    + c^{-4} \bigg({-}a_i (\vect a \cdot \vect x)
      - \frac{2c^2}{3} (\vect a \cdot \vect x) R_{0i0j} x^j
    \nonumber\\
  &\qquad- \frac{c^2}{3} a_i R_{0l0m} x^l x^m \bigg)
    + c^{-6} a_i (\vect a \cdot \vect x)^2
    + \Or(c^{-2}x^4) + \Or(c^{-3}x^3),\\
  \Gamma^s_{ij}
  &= c^{-2} \bigg(\frac{c^2}{3} \left\{2 R_{0(ij)k}
        + \frac{1}{4} (5 R_{0(ij)k;l} - R_{0kl(i;j)}) x^l \right\} x^k
    \nonumber\\
  &\qquad\textcolor{blue}{{}+ \frac{c^2}{20} (3R_{0(ij)l;mn}
        - R_{0lm(i;j)n}) x^l x^m x^n}
      \bigg) \nonumber\\
  &\quad+ c^{-4} \left(-\frac{2c^2}{3} (\vect a \cdot \vect x)
        R_{0(ij)l} x^l\right)
    + \Or(c^{-2}x^4) + \Or(c^{-3}x^3),\\
  \Gamma^i_{ss}
  &= c^{-2} \bigg(a^i + c^2 \tensor{R}{_0^i_{0j}} x^j
      + (\vect\eta \times \vect x)^i
      + (\vect\omega \times (\vect\omega \times \vect x))^i
      + \frac{c^2}{2} (2\tensor{R}{_0^i_{0l;m}}
        - \tensor{R}{_{0l0m}^{;i}}) x^l x^m \nonumber\\
  &\qquad\textcolor{blue}{{}+ \frac{c^2}{6}
        (2\tensor{R}{_0^i_{0l;mn}} - \tensor{R}{_{0l0m;n}^i}) x^l x^m x^n}
      \bigg)
    + c^3 (- 2 c (\vect\omega \times \vect x)^k
        \tensor{R}{_{0j}^i_k} x^j) \nonumber\\
  &\quad+ c^{-4} \bigg((\vect a \cdot \vect x) a^i
      + 2 c^{2} (\vect a \cdot \vect x) \tensor{R}{_0^i_{0j}} x^j
      - \frac{c^2}{3} a^j \tensor{R}{^i_{ljm}} x^l x^m \nonumber\\
  &\qquad- (\vect\omega \times \vect x)^i (\vect b \cdot \vect x
        + 2 \vect a \cdot (\vect\omega \times \vect x)) \bigg)
    + \Or(c^{-2}x^4) + \Or(c^{-3}x^3),\\
  \Gamma^i_{sj}
  &= - c^{-1} \tensor{\varepsilon}{^i_{jk}} \omega^k
    + c^{-2} \bigg({-}c^2 \tensor{R}{_{0k}^i_j} x^k
      + c^2 \left\{\frac{1}{6} \tensor{R}{_{0j}^i_{l;m}}
        - \frac{1}{2} \tensor{R}{_{0l}^i_{j;m}}
        - \frac{1}{6} \tensor{R}{_{0l}^i_{m;j}} \right\} x^l x^m
    \nonumber\\
  &\qquad\textcolor{blue}{{}+ \frac{c^2}{12}
        (\tensor{R}{_{0j}^i_{l;mn}} - 2 \tensor{R}{_{0l}^i_{j;mn}}
          - \tensor{R}{_{0l}^i_{m;nj}})
        x^l x^m x^n} \bigg) \nonumber\\
  &\quad+ c^{-3} \left(-(\vect\omega \times \vect x)^i a_j
      - \frac{c^2}{3} (\vect\omega \times \vect x)^l
        (\tensor{R}{_{lk}^i_j} + \tensor{R}{_l^i_{kj}}) x^k
      - c^2 (\vect\omega \times \vect x)^i R_{0j0k} x^k \right)
    \nonumber\\
  &\quad+ c^{-4} \left(- \frac{c^2}{3} (\vect a \cdot \vect x)
        (\tensor{R}{_{0k}^i_j} + \tensor{R}{_0^i_{kj}}) x^k
      + \frac{c^2}{3} a_j \tensor{R}{_{0l}^i_m} x^l x^m \right)
    + c^{-5} a_j (\vect a \cdot \vect x)
        (\vect\omega \times \vect x)^i \nonumber\\
  &\quad+ \Or(c^{-2}x^4) + \Or(c^{-3}x^3),\\
  \Gamma^i_{jk}
  &= c^{-2} \bigg(- \frac{c^2}{3} \left\{2 \tensor{R}{^i_{(jk)l}}
        + \frac{1}{4} (5 \tensor{R}{^i_{(jk)l;m}}
          - \tensor{R}{^i_{lm(j;k)}}) x^m \right\} x^l \nonumber\\
  &\qquad\textcolor{blue}{{}- \frac{c^2}{20} (3\tensor{R}{^i_{(jk)l;mn}}
        - \tensor{R}{^i_{lm(j;k)n}}) x^l x^m x^n} \bigg)\nonumber\\
  &\quad+ c^{-3} (2 c^2 (\vect\omega \times \vect x)^i R_{0(jk)l} x^l)
    + \Or(c^{-2}x^4) + \Or(c^{-3}x^3).
\end{align}
\end{subequations}
Note that, according to \eqref{eq:curvature_c_order}, we have treated
the curvature tensor as being of order $c^{-2}$.

Using the above Christoffel symbols, one can compute the parallely
transported frame \eqref{eq:frame} to higher order of expansion, which
reads
\begin{subequations}
\begin{align}
  (\e_0)^s &= 1 + c^{-2} \left(-\vect a \cdot \vect x
               - \frac{c^2}{2} R_{0l0m} x^l x^m
               - \frac{c^2}{6} R_{0l0m;n} x^l x^m x^n
               \textcolor{blue}{{}- \frac{c^2}{24} R_{0k0l;mn} x^k x^l
                 x^m x^n} \right) \nonumber\\
           &\quad+ c^{-4} \left((\vect a \cdot \vect x)^2
               + \frac{5c^2}{6} (\vect a \cdot \vect x)
                 R_{0l0m} x^l x^m \right)
             - c^{-6} (\vect a \cdot \vect x)^3
             + \Or(c^{-2}x^5) + \Or(c^{-3}x^4),\\
  (\e_0)^i &= - c^{-1} (\vect\omega \times \vect x)^i
             + c^{-2} \left(\frac{c^2}{2} \tensor{R}{_{0l}^i_m}
                 x^l x^m
               + \frac{c^2}{6} \tensor{R}{_{0l}^i_{m;n}} x^l x^m x^n
             \textcolor{blue}{{}+ \frac{c^2}{24}
               \tensor{R}{_{0k}^i_{l;mn}} x^k x^l x^m x^n} \right)\nonumber\\
           &\quad+ c^{-3} \left((\vect a \cdot \vect x)
                 (\vect\omega \times\vect x)^i
               + \frac{c^2}{2} (\vect\omega \times \vect x)^i
                 R_{0l0m} x^l x^m \right)
             + c^{-4} \left(-\frac{c^2}{3} (\vect a \cdot \vect x)
                 \tensor{R}{_{0l}^i_m} x^l x^m \right) \nonumber\\
           &\quad- c^{-5} (\vect\omega \times \vect x)^i
                 (\vect a \cdot \vect x)^2
             + \Or(c^{-2}x^5) + \Or(c^{-3}x^4),\\
  (\e_i)^s &= c^{-2} \left(-\frac{c^2}{6} R_{0lim} x^l x^m
               - \frac{c^2}{12} R_{0lim;n} x^l x^m x^n
               \textcolor{blue}{{}- \frac{c^2}{40} R_{0kil;mn} x^k x^l
                 x^m x^n} \right) \nonumber\\
           &\quad+ c^{-4} \left(\frac{c^2}{6} (\vect a \cdot \vect x)
                 R_{0lim} x^l x^m \right)
             + \Or(c^{-2}x^5) + \Or(c^{-3}x^4),\\
  (\e_i)^j &= \delta^j_i
             + c^{-2} \left(\frac{c^2}{6} \tensor{R}{^j_{lim}} x^l x^m
               + \frac{c^2}{12} \tensor{R}{^j_{lim;n}} x^l x^m x^n
               \textcolor{blue}{{}+ \frac{c^2}{40}
                 \tensor{R}{^j_{kil;mn}} x^k x^l x^m x^n} \right)
             \nonumber\\
           &\quad+ c^{-3} \left(\frac{c^2}{6}
                 (\vect\omega \times \vect x)^j R_{0lim} x^l x^m
             \right)
             + \Or(c^{-2}x^5) + \Or(c^{-3}x^4).
\end{align}
\end{subequations}
For the dual frame, we also obtain that the $x$ dependence starts at
order $c^{-2}$:
\begin{subequations}
\begin{align}
  (\e^0)_s &= 1 + c^{-2} \left(\vect a \cdot \vect x
               + \frac{c^2}{2} R_{0l0m} x^l x^m \right)
             + \Or(c^{-2}x^3)\\
  (\e^0)_i &= c^{-2} \left(\frac{c^2}{6} R_{0lim} x^l x^m \right)
             + \Or(c^{-2}x^3)\\
  (\e^i)_s &= c^{-1} (\vect\omega \times \vect x)^i
             + c^{-2} \left(-\frac{c^2}{2} \tensor{R}{^i_{l0m}}
                 x^l x^m \right)
             + \Or(c^{-2}x^3)\\
  (\e^i)_j &= \delta^i_j
             + c^{-2} \left(-\frac{c^2}{6}
                 \tensor{R}{^i_{ljm}} x^l x^m \right)
            + \Or(c^{-2}x^3)
\end{align}
\end{subequations}

From this, we can compute the higher-order corrections to the
connection form, the nontrivial components of which read
\begin{subequations}
\begin{align}
  \tensor{\omega}{_s^0_i}
  &= c^{-2} \left(a_i + c^2 R_{0i0l} x^l
      + \frac{c^2}{2} R_{0i0l;m} x^l x^m
      \textcolor{blue}{{}+ \frac{c^2}{6} R_{0i0l;mn} x^l x^m x^n}
    \right) \nonumber\\
  &\quad+ c^{-3} \left(\frac{c^2}{2} (\vect\omega \times \vect x)^k
        R_{0ikl} x^l \right)
    + c^{-4} \left(\frac{c^2}{2} (\vect a \cdot \vect x) R_{0i0l} x^l
    \right)
    + \Or(c^{-2}x^4) + \Or(c^{-3}x^3), \displaybreak[0]\\
  \tensor{\omega}{_i^0_j}
  &= c^{-2} \left(\frac{c^2}{2} R_{0jil} x^l
      + \frac{c^2}{3} R_{0jil;m} x^l x^m
      \textcolor{blue}{{}+ \frac{c^2}{8} R_{0jil;mn} x^l x^m x^n}
    \right) + \Or(c^{-2}x^4) + \Or(c^{-3}x^3),\\
  \tensor{\omega}{_s^i_j}
  &= - c^{-1} \tensor{\varepsilon}{^i_{jk}} \omega^k
    + c^{-2} \left(-c^2 \tensor{R}{^i_{j0l}} x^l
      - \frac{c^2}{2} \tensor{R}{^i_{j0l;m}} x^l x^m
      \textcolor{blue}{{}- \frac{c^2}{6} \tensor{R}{^i_{j0l;mn}} x^l
        x^m x^n} \right)
    \nonumber\\
  &\quad+ c^{-3} \left(-\frac{c^2}{2} (\vect\omega \times \vect x)^k
        \tensor{R}{^i_{jkl}} x^l \right)
    + c^{-4} \left(-\frac{c^2}{2} (\vect a \cdot \vect x)
        \tensor{R}{^i_{j0l}} x^l \right)
    + \Or(c^{-2}x^4) + \Or(c^{-3}x^3),\\
  \tensor{\omega}{_k^i_j}
  &= c^{-2} \left(-\frac{c^2}{2} \tensor{R}{^i_{jkl}} x^l
      - \frac{c^2}{3} \tensor{R}{^i_{jkl;m}} x^l x^m
      \textcolor{blue}{{}- \frac{c^2}{8} \tensor{R}{^i_{jkl;mn}} x^l
        x^m x^n} \right)
    + \Or(c^{-2}x^4) + \Or(c^{-3}x^3).
\end{align}
\end{subequations}
The component of the inverse metric that is needed for the computation
of the Dirac Hamiltonian takes the following form including the newly
computed higher-order corrections:
\begin{align}
  g^{ss} &= -1 + c^{-2} \left(2 \vect a \cdot \vect x
             + c^2 R_{0l0m} x^l x^m
             + \frac{c^2}{3} R_{0l0m;n} x^l x^m x^n
             \textcolor{blue}{{}+ \frac{c^2}{12} R_{0k0l;mn} x^k x^l
               x^m x^n} \right) \nonumber\\
         &\quad+ c^{-4} \left(-3 (\vect a \cdot \vect x)^2
             - \frac{8c^2}{3} (\vect a \cdot \vect x) R_{0l0m} x^l x^m
           \right)
           + c^{-6}4(\vect a \cdot \vect x)^3
           + \Or(c^{-2}x^5) + \Or(c^{-3}x^4)
\end{align}

Using all these ingredients, we can finally compute the Dirac
Hamiltonian in our coordinates and frame to the necessary order (as
for the original Dirac Hamiltonian \eqref{eq:Dirac_FNC}, the
computation is rather tedious, but straightforward):
\begin{align} \label{eq:Dirac_FNC_higher_order}
  H_\text{Dirac}
  &= \gamma^0 \left\{mc^2 + c^0 \left(m \vect a \cdot \vect x
        + \frac{m c^2}{2} R_{0l0m} x^l x^m
        \textcolor{blue}{{}+ \frac{m c^2}{6} R_{0l0m;n} x^l x^m x^n}
        \right) + \Or(c^0x^4) \right\} \nonumber\\
  &\quad- \gamma^i \left\{c^0 \left(\frac{m c^2}{6} R_{0lim} x^l x^m
        \textcolor{blue}{{}+ \frac{m c^2}{12} R_{0lim;n} x^l x^m x^n}
        \right) + \Or(c^0x^4) \right\} \nonumber\\
  &\quad+\mathbb{1} \bigg{\{}{-} q A_\tau
      + \I (\vect\omega \times \vect x)^i D_i
      + c^{-1} \bigg({-}\frac{\I c^2}{2} \tensor{R}{_{0l}^i_m} x^l x^m
          D_i
        + \frac{\I c^2}{12} R_{0l;m} x^l x^m \nonumber\\
  &\qquad\quad- \frac{\I c^2}{6} \tensor{R}{_{0l}^i_{m;n}} x^l x^m x^n
          D_i
        \textcolor{blue}{{}+ \frac{\I c^2}{24} R_{0l;mn} x^l x^m x^n
        - \frac{\I c^2}{24} \tensor{R}{_{0k}^i_{l;mn}} x^k x^l x^m x^n
          D_i} \bigg) \nonumber\\
  &\qquad+ c^{-3} \left( \frac{\I c^2}{4} (\vect a \cdot \vect x)
          R_{0l} x^l
      - \frac{\I c^2}{6} (\vect a \cdot \vect x) \tensor{R}{_{0l}^i_m}
        x^l x^m D_i \right) \bigg\} \nonumber\\
  &\quad- \gamma^0 \gamma^j \bigg\{c \I D_j
      + c^{-1} \bigg(\frac{\I}{2} a_j
        + \I (\vect a \cdot \vect x) D_j
        + \frac{\I c^2}{4} (R_{0j0l} - R_{jl}) x^l
        + \frac{\I c^2}{2} R_{0l0m} x^l x^m D_j \nonumber\\
  &\qquad\quad+ \frac{\I c^2}{6} \tensor{R}{^i_{ljm}} x^l x^m D_i
        + \frac{\I c^2}{12} (R_{0j0l;m} - 2 R_{jl;m}) x^l x^m
        + \frac{\I c^2}{6} R_{0l0m;n} x^l x^m x^n D_j \nonumber\\
  &\qquad\quad+ \frac{\I c^2}{12} \tensor{R}{^i_{ljm;n}} x^l x^m x^n D_i
        \textcolor{blue}{{}
        + \frac{\I c^2}{48} (R_{0j0l;mn} - 3 R_{jl;mn}) x^l x^m x^n}
    \nonumber\\
  &\qquad\quad\textcolor{blue}{{}
        + \frac{\I c^2}{24} R_{0k0l;mn} x^k x^l x^m x^n D_j
        + \frac{\I c^2}{40} \tensor{R}{^i_{kjl;mn}} x^k x^l x^m x^n
          D_i} \bigg) \nonumber\\
  &\qquad+ c^{-3} \left( -\frac{\I c^2}{4} (\vect a \cdot \vect x)
          R_{jl} x^l
        + \frac{\I c^2}{6} (\vect a \cdot \vect x) R_{0l0m} x^l x^m
          D_j
        + \frac{\I c^2}{6} (\vect a \cdot \vect x)
          \tensor{R}{^i_{ljm}} x^l x^m D_i \right) \bigg\} \nonumber\\
  &\quad+ \gamma^i \gamma^j \bigg\{
      {-} \frac{\I}{4} \varepsilon_{ijk} \omega^k
      + c^{-1} \bigg(\frac{\I c^2}{4} R_{0ijl} x^l
        + \frac{\I c^2}{6} R_{0lim} x^l x^m D_j
        + \frac{\I c^2}{12} R_{0ijl;m} x^l x^m \nonumber\\
  &\qquad\quad+ \frac{\I c^2}{12} R_{0lim;n} x^l x^m x^n D_j
        \textcolor{blue}{{}+ \frac{\I c^2}{48} R_{0ijl;mn} x^l x^m x^n
        + \frac{\I c^2}{40} R_{0kil;mn} x^k x^l x^m x^n D_j} \bigg)
    \nonumber\\
  &\qquad+ c^{-3} \frac{\I c^2}{6} (\vect a \cdot \vect x) R_{0lim}
        x^l x^m D_j \bigg\}
    + \Or(c^{-1}x^4) + \Or(c^{-2}x^3)
\end{align}

\section{Details of the post-Newtonian expansion}
\label{sec:app_details_PN}

The equations that arise from the Dirac equation when inserting the
post-Newtonian ansatz \eqref{eq:Dirac_PN_ansatz} are
\begin{subequations} \label{eq:Dirac_expanded}
  \begin{align} \label{eq:Dirac_expanded_1}
    &\bigg\{\I D_\tau - m \vect a \cdot \vect x
      - \frac{mc^2}{2} R_{0l0m} x^l x^m
      - \frac{m c^2}{6} R_{0l0m;n} x^l x^m x^n
      - \I(\vect\omega \times \vect x)^i D_i
      + \frac{1}{2} \vect\sigma \cdot \vect\omega \nonumber\\
    &\quad+ c^{-1} \bigg(
        \frac{\I c^2}{2} \tensor{R}{_{0l}^i_m} x^l x^m D_i
        - \frac{\I c^2}{6} R_{0l;m} x^l x^m
        - \frac{c^2}{12} \tensor{\varepsilon}{^{ij}_k} \sigma^k
          R_{0ijl;m} x^l x^m
        + \frac{\I c^2}{6} \tensor{R}{_{0l}^i_{m;n}} x^l x^m x^n D_i
      \nonumber\\
    &\qquad- \frac{\I c^2}{16} R_{0l;mn} x^l x^m x^n
        - \frac{c^2}{48} \tensor{\varepsilon}{^{ij}_k} \sigma^k
          R_{0ijl;mn} x^l x^m x^n
        + \frac{\I c^2}{24} \tensor{R}{_{0k}^i_{l;mn}} x^k x^l x^m x^n
          D_i
        + \sigma^i \sigma^j \bigg[
          \frac{\I c^2}{4} R_{0ijl} x^l \nonumber\\
    &\qquad\quad+ \frac{\I c^2}{6} R_{0lim} x^l x^m D_j
          + \frac{\I c^2}{12} R_{0lim;n} x^l x^m x^n D_j
          + \frac{\I c^2}{40} R_{0kil;mn} x^k x^l x^m x^n D_j
        \bigg] \bigg) \nonumber\\
    &\quad+ c^{-3} \left( -\frac{\I c^2}{4} (\vect a \cdot \vect x)
          R_{0l} x^l
        + \frac{\I c^2}{6} (\vect a \cdot \vect x)
          \tensor{R}{_{0l}^i_m} x^l x^m D_i
        + \frac{\I c^2}{6} (\vect a \cdot \vect x) \sigma^i \sigma^j
          R_{0lim} x^l x^m D_j \right) \nonumber\\
    &\quad+ \Or(c^0x^4) + \Or(c^{-2}x^3) \bigg\}
      \tilde{\psi}_A \nonumber\\
    &= -\sigma^j \bigg\{\I c D_j
      + \frac{mc^2}{6} R_{0ljm} x^l x^m
      + \frac{m c^2}{12} R_{0ljm;n} x^l x^m x^n
      + c^{-1} \bigg(\frac{\I}{2} a_j
        + \I (\vect a \cdot \vect x) D_j
        + \frac{\I c^2}{4} (R_{0j0l} - R_{jl}) x^l \nonumber\\
    &\qquad + \frac{\I c^2}{2} R_{0l0m} x^l x^m D_j
        + \frac{\I c^2}{6} \tensor{R}{^i_{ljm}} x^l x^m D_i
        + \frac{\I c^2}{12} (R_{0j0l;m} - 2 R_{jl;m}) x^l x^m
        + \frac{\I c^2}{6} R_{0l0m;n} x^l x^m x^n D_j \nonumber\\
    &\qquad + \frac{\I c^2}{12} \tensor{R}{^i_{ljm;n}} x^l x^m x^n D_i
        + \frac{\I c^2}{48} (R_{0j0l;mn} - 3 R_{jl;mn}) x^l x^m x^n
        + \frac{\I c^2}{24} R_{0k0l;mn} x^k x^l x^m x^n D_j
    \nonumber\\
    &\qquad+ \frac{\I c^2}{40} \tensor{R}{^i_{kjl;mn}} x^k x^l x^m x^n
        D_i \bigg)
      + c^{-3} \bigg( -\frac{\I c^2}{4} (\vect a \cdot \vect x) R_{jl}
        x^l 
      + \frac{\I c^2}{6} (\vect a \cdot \vect x) R_{0l0m} x^l x^m D_j
      \nonumber\\
    &\qquad+ \frac{\I c^2}{6} (\vect a \cdot \vect x)
        \tensor{R}{^i_{ljm}} x^l x^m D_i \bigg)
      + \Or(c^0x^4) + \Or(c^{-2}x^3) \bigg\}
      \tilde{\psi}_B \; , \displaybreak[0]\\
    \label{eq:Dirac_expanded_2}
    &\bigg\{2 mc^2 + \I D_\tau + m \vect a \cdot \vect x
      + \frac{mc^2}{2} R_{0l0m} x^l x^m
      + \frac{m c^2}{6} R_{0l0m;n} x^l x^m x^n
      - \I(\vect\omega \times \vect x)^i D_i
      + \frac{1}{2} \vect\sigma \cdot \vect\omega \nonumber\\
    &\quad+ c^{-1} \bigg(
        \frac{\I c}{2} \tensor{R}{_{0l}^i_m} x^l x^m D_i
        - \frac{\I c^2}{6} R_{0l;m} x^l x^m
        - \frac{c^2}{12} \tensor{\varepsilon}{^{ij}_k} \sigma^k
          R_{0ijl;m} x^l x^m
        + \frac{\I c^2}{6} \tensor{R}{_{0l}^i_{m;n}} x^l x^m x^n D_i
      \nonumber\\
    &\qquad- \frac{\I c^2}{16} R_{0l;mn} x^l x^m x^n
        - \frac{c^2}{48} \tensor{\varepsilon}{^{ij}_k} \sigma^k
          R_{0ijl;mn} x^l x^m x^n
        + \frac{\I c^2}{24} \tensor{R}{_{0k}^i_{l;mn}} x^k x^l x^m x^n
          D_i
        + \sigma^i \sigma^j \bigg[
          \frac{\I c^2}{4} R_{0ijl} x^l \nonumber\\
    &\qquad\quad+ \frac{\I c^2}{6} R_{0lim} x^l x^m D_j
          + \frac{\I c^2}{12} R_{0lim;n} x^l x^m x^n D_j
          + \frac{\I c^2}{40} R_{0kil;mn} x^k x^l x^m x^n D_j
        \bigg] \bigg) \nonumber\\
    &\quad+ c^{-3} \left( -\frac{\I c^2}{4} (\vect a \cdot \vect x)
          R_{0l} x^l
        + \frac{\I c^2}{6} (\vect a \cdot \vect x)
          \tensor{R}{_{0l}^i_m} x^l x^m D_i
        + \frac{\I c^2}{6} (\vect a \cdot \vect x) \sigma^i \sigma^j
          R_{0lim} x^l x^m D_j \right) \nonumber\\
    &\quad+ \Or(c^0x^4) + \Or(c^{-2}x^3) \bigg\}
      \tilde{\psi}_B \nonumber\\
    &= -\sigma^j \bigg\{\I c D_j
      - \frac{mc^2}{6} R_{0ljm} x^l x^m
      - \frac{m c^2}{12} R_{0ljm;n} x^l x^m x^n
      + c^{-1} \bigg(\frac{\I}{2} a_j
        + \I (\vect a \cdot \vect x) D_j
        + \frac{\I c^2}{4} (R_{0j0l} - R_{jl}) x^l \nonumber\\
    &\qquad + \frac{\I c^2}{2} R_{0l0m} x^l x^m D_j
        + \frac{\I c^2}{6} \tensor{R}{^i_{ljm}} x^l x^m D_i
        + \frac{\I c^2}{12} (R_{0j0l;m} - 2 R_{jl;m}) x^l x^m
        + \frac{\I c^2}{6} R_{0l0m;n} x^l x^m x^n D_j \nonumber\\
    &\qquad + \frac{\I c^2}{12} \tensor{R}{^i_{ljm;n}} x^l x^m x^n D_i
        + \frac{\I c^2}{48} (R_{0j0l;mn} - 3 R_{jl;mn}) x^l x^m x^n
        + \frac{\I c^2}{24} R_{0k0l;mn} x^k x^l x^m x^n D_j
    \nonumber\\
    &\qquad+ \frac{\I c^2}{40} \tensor{R}{^i_{kjl;mn}} x^k x^l x^m x^n
        D_i \bigg)
      + c^{-3} \bigg( -\frac{\I c^2}{4} (\vect a \cdot \vect x) R_{jl}
        x^l 
      + \frac{\I c^2}{6} (\vect a \cdot \vect x) R_{0l0m} x^l x^m D_j
      \nonumber\\
    &\qquad+ \frac{\I c^2}{6} (\vect a \cdot \vect x)
        \tensor{R}{^i_{ljm}} x^l x^m D_i \bigg)
      + \Or(c^0x^4) + \Or(c^{-2}x^3) \bigg\}
      \tilde{\psi}_A \; ,
  \end{align}
\end{subequations}
where $D_\tau = \partial_\tau - \I q A_\tau, D_i = \partial_i - \I q
A_i$ denotes the electromagnetic covariant derivative.  Note that we
used the Pauli matrix identity $\sigma^i \sigma^j = \delta^{ij}
\mathbb{1} + \I \tensor{\varepsilon}{^{ij}_k} \sigma^k$ for the
simplifications $\sigma^i \sigma^j \varepsilon_{ijk} \omega^k = 2 \I
\vect\sigma \cdot \vect\omega$ and $\sigma^i \sigma^j R_{0ijl;m} =
-R_{0l;m} + \I \tensor{\varepsilon}{^{ij}_k} \sigma^k R_{0ijl;m}$.

At order $c^{-1}$, \eqref{eq:Dirac_expanded_1} yields
\begin{align}
  &\bigg\{\I D_\tau - m \vect a \cdot \vect x
      - \frac{mc^2}{2} R_{0l0m} x^l x^m
      - \I (\vect\omega \times \vect x)^i D_i
      + \frac{1}{2} \vect\sigma \cdot \vect\omega
      + \Or(x^3) \bigg\} \tilde{\psi}_A^{(1)} \nonumber\\
  &+ \bigg\{\frac{\I c^2}{2} \tensor{R}{_{0l}^i_m} x^l x^m D_i
      - \frac{\I c^2}{6} R_{0l;m} x^l x^m
      - \frac{c^2}{12} \tensor{\varepsilon}{^{ij}_k} \sigma^k
        R_{0ijl;m} x^l x^m
      + \frac{\I c^2}{6} \tensor{R}{_{0l}^i_{m;n}} x^l x^m x^n D_i
    \nonumber\\
  &\quad + \frac{\I c^2}{4} \sigma^i \sigma^j R_{0ijl} x^l
      + \frac{\I c^2}{6} \sigma^i \sigma^j R_{0lim} x^l x^m D_j
      + \frac{\I c^2}{12} \sigma^i \sigma^j R_{0lim;n} x^l x^m x^n D_j
      + \Or(x^3) \bigg\} \tilde{\psi}_A^{(0)} \nonumber\\
  &= - \I \sigma^j D_j \tilde{\psi}_B^{(2)}
    + \left\{- \frac{mc^2}{6} \sigma^i R_{0lim} x^l x^m
      - \frac{m c^2}{12} \sigma^i R_{0lim;n} x^l x^m x^n
      + \Or(x^4) \right\} \tilde{\psi}_B^{(1)} .
\end{align}
Using \eqref{eq:PN_elim_psi_B_1} and \eqref{eq:PN_elim_psi_B_2} to
eliminate the $\tilde{\psi}_B$, this may be rewritten as
\begin{align} \label{eq:PN_psi_A_1}
  &\bigg\{\I D_\tau - m \vect a \cdot \vect x
      - \frac{mc^2}{2} R_{0l0m} x^l x^m
      - \I (\vect\omega \times \vect x)^i D_i
      + \frac{1}{2} \vect\sigma \cdot \vect\omega
      + \Or(x^3) \bigg\} \tilde{\psi}_A^{(1)} \nonumber\\
  &+ \bigg\{\frac{\I c^2}{2} \tensor{R}{_{0l}^i_m} x^l x^m D_i
      - \frac{\I c^2}{6} R_{0l;m} x^l x^m
      - \frac{c^2}{12} \tensor{\varepsilon}{^{ij}_k} \sigma^k
        R_{0ijl;m} x^l x^m
      + \frac{\I c^2}{6} \tensor{R}{_{0l}^i_{m;n}} x^l x^m x^n D_i
    \nonumber\\
  &\quad + \frac{\I c^2}{4} \sigma^i \sigma^j R_{0ijl} x^l
      + \frac{\I c^2}{6} \sigma^i \sigma^j R_{0lim} x^l x^m D_j
      + \frac{\I c^2}{12} \sigma^i \sigma^j R_{0lim;n} x^l x^m x^n D_j
      + \Or(x^3) \bigg\} \tilde{\psi}_A^{(0)} \nonumber\\
  &= - \frac{1}{2m} (\vect\sigma \cdot \vect D)^2 \tilde{\psi}_A^{(1)}
    + \bigg\{- \frac{\I c^2}{12} \sigma^j \sigma^i R_{0lim} D_j
        (x^l x^m \cdot)
      - \frac{\I c^2}{24} \sigma^j \sigma^i R_{0lim;n} D_j (x^l x^m
        x^n \cdot) \nonumber\\ 
  &\qquad+ \frac{\I c^2}{12} \sigma^i \sigma^j R_{0lim} x^l x^m D_j
      + \frac{\I c^2}{24} \sigma^i \sigma^j R_{0lim;n} x^l x^m x^n D_j
      + \Or(x^3)\bigg\} \tilde{\psi}_A^{(0)}
\end{align}
From this, we can read off the next-to-leading-order Hamiltonian
$H^{(1)}$ according to \eqref{eq:Pauli_PN_1}, giving
\begin{align}
  H^{(1)}
  &= - \frac{\I c^2}{2} \tensor{R}{_{0l}^i_m} x^l x^m D_i
    + \frac{\I c^2}{6} R_{0l;m} x^l x^m
    + \frac{c^2}{12} \tensor{\varepsilon}{^{ij}_k} \sigma^k R_{0ijl;m}
      x^l x^m
    - \frac{\I c^2}{6} \tensor{R}{_{0l}^i_{m;n}} x^l x^m x^n D_i
    \nonumber\\
  &\quad - \frac{\I c^2}{4} \sigma^i \sigma^j R_{0ijl} x^l
    - \frac{\I c^2}{12} R_{0lim}
      (\sigma^i \sigma^j x^l x^m D_j
      + \sigma^j \sigma^i D_j(x^l x^m \cdot)) \nonumber\\
  &\quad- \frac{\I c^2}{24} R_{0lim;n}
      (\sigma^i \sigma^j x^l x^m x^n D_j
      + \sigma^j \sigma^i D_j (x^l x^m x^n \cdot))
    + \Or(x^3) \nonumber\\
  &= \frac{\I c^2}{3} R_{0l} x^l
    - \frac{c^2}{4} \tensor{\varepsilon}{^{ij}_k} \sigma^k R_{0lij}
      x^l
    - \frac{2\I c^2}{3} \tensor{R}{_{0l}^j_m} x^l x^m D_j
    + \frac{\I c^2}{24}(5 R_{0l;m} - \tensor{R}{_{0l}^i_{m;i}}) x^l
      x^m \nonumber\\
  &\quad- \frac{c^2}{8} \tensor{\varepsilon}{^{ij}_k} \sigma^k
      R_{0lij;m} x^l x^m
    - \frac{\I c^2}{4} \tensor{R}{_{0l}^j_{m;n}} x^l x^m x^n D_j
    + \Or(x^3),
\end{align}
where we again used $\sigma^i \sigma^j = \delta^{ij} \mathbb{1} + \I
\tensor{\varepsilon}{^{ij}_k} \sigma^k$ for simplifications, as well
as the Bianchi identities.  The difference of this result to the
corresponding one in the master's thesis \cite{Alibabaei:2022Thesis}
on which the present article is based, arising from oversights in
\cite{Alibabaei:2022Thesis} regarding the consistent calculation of
the order $x^2$ terms, consists solely in the appearance of the terms
containing covariant derivatives of the curvature tensor.  Note that
$H^{(0)}$ read off from \eqref{eq:PN_psi_A_1} is the same as the one
calculated above in \eqref{eq:Pauli_PN_Ham_0}.

\eqref{eq:Dirac_expanded_2} at order $c^{-1}$ gives the following:
\begin{align} \label{eq:PN_elim_psi_B_3}
  &2m \tilde{\psi}_B^{(3)}
    + \bigg\{\I D_\tau + m \vect a \cdot \vect x
      + \frac{mc^2}{2} R_{0l0m} x^l x^m
      + \frac{m c^2}{6} R_{0l0m;n} x^l x^m x^n
      - \I (\vect\omega \times \vect x)^i D_i
      + \frac{1}{2} \vect\sigma \cdot \vect\omega \nonumber\\
  &\quad + \Or(x^4) \bigg\} \tilde{\psi}_B^{(1)} \nonumber\\
  &= - \I \sigma^j D_j \tilde{\psi}_A^{(2)}
    + \left\{\frac{mc^2}{6} \sigma^i R_{0lim} x^l x^m
      + \frac{m c^2}{12} \sigma^i R_{0lim;n} x^l x^m x^n
      + \Or(x^4)
    \right\} \tilde{\psi}_A^{(1)} \nonumber\\
  &\quad - \sigma^j \bigg\{ \frac{\I}{2} a_j
      + \I (\vect a \cdot \vect x) D_j
      + \frac{\I c^2}{4} (R_{0j0l} - R_{jl}) x^l
      + \frac{\I c^2}{2} R_{0l0m} x^l x^m D_j
      + \frac{\I c^2}{6} \tensor{R}{^i_{ljm}} x^l x^m D_i \nonumber\\
  &\qquad + \frac{\I c^2}{12} (R_{0j0l;m} - 2 R_{jl;m}) x^l x^m
      + \frac{\I c^2}{6} R_{0l0m;n} x^l x^m x^n D_j
      + \frac{\I c^2}{12} \tensor{R}{^i_{ljm;n}} x^l x^m x^n D_i
    \nonumber\\
  &\qquad + \frac{\I c^2}{48} (R_{0j0l;mn} - 3 R_{jl;mn}) x^l x^m x^n
      + \frac{\I c^2}{24} R_{0k0l;mn} x^k x^l x^m x^n D_j
      + \frac{\I c^2}{40} \tensor{R}{^i_{kjl;mn}} x^k x^l x^m x^n D_i
    \nonumber\\
  &\qquad + \Or(x^4) \bigg\} \tilde{\psi}_A^{(0)}
\end{align}
With \eqref{eq:PN_elim_psi_B_1} to eliminate $\tilde{\psi}_B^{(1)}$,
this can be used to express $\tilde{\psi}_B^{(3)}$ in terms of the
$\tilde{\psi}_A$.  Note however that this will involve the term
$-\frac{\I D_\tau}{2m} \tilde{\psi}_B^{(1)} = -\frac{\I D_\tau}{4m^2}
(-\I \vect\sigma \cdot \vect D) \tilde{\psi}_A^{(0)}$, such that we
need to re-use the Pauli equation \eqref{eq:Pauli_PN_0} for
$\tilde{\psi}_A^{(0)}$ to fully eliminate the time derivative in the
resulting expression.  Explicitly, the term in question evaluates to
\begin{align} \label{eq:PN_elim_psi_B_3_time_der}
  \frac{-\I D_\tau}{4m^2} (-\I \vect\sigma \cdot \vect D)
    \tilde{\psi}_A^{(0)}
  &= - \frac{1}{4m^2} \{
      [D_\tau, \vect\sigma \cdot \vect D]
      + (-\I \vect\sigma \cdot \vect D) \I D_\tau \}
    \tilde{\psi}_A^{(0)} \nonumber\\
  &= - \frac{1}{4m^2} \{ \I q \vect\sigma \cdot \vect E
      + (-\I \vect\sigma \cdot \vect D) (H^{(0)} + q A_\tau) \}
    \tilde{\psi}_A^{(0)} \; ,
\end{align}
where $E_i = \partial_i A_\tau - \partial_\tau A_i$ is the electric
field (note that up to higher-order corrections, these are indeed the
electric field components in an orthonormal basis).

We finally need the next order of expansion in $c^{-1}$ in order to
compute the Hamiltonian at order $c^{-2}$.
\eqref{eq:Dirac_expanded_1} at order $c^{-2}$ is
\begin{align} \label{eq:PN_psi_A_2}
  &\bigg\{\I D_\tau - m \vect a \cdot \vect x
      - \frac{mc^2}{2} R_{0l0m} x^l x^m
      - \I (\vect\omega \times \vect x)^i D_i
      + \frac{1}{2} \vect\sigma \cdot \vect\omega
      + \Or(x^3)
    \bigg\} \tilde{\psi}_A^{(2)} \nonumber\\
  &+ \bigg\{\frac{\I c^2}{2} \tensor{R}{_{0l}^i_m} x^l x^m D_i
      - \frac{\I c^2}{6} R_{0l;m} x^l x^m
      - \frac{c^2}{12} \tensor{\varepsilon}{^{ij}_k} \sigma^k
        R_{0ijl;m} x^l x^m
      + \frac{\I c^2}{6} \tensor{R}{_{0l}^i_{m;n}} x^l x^m x^n D_i
    \nonumber\\
  &\quad + \frac{\I c^2}{4} \sigma^i \sigma^j R_{0ijl} x^l
      + \frac{\I c^2}{6} \sigma^i \sigma^j R_{0lim} x^l x^m D_j
      + \frac{\I c^2}{12} \sigma^i \sigma^j R_{0lim;n} x^l x^m x^n D_j
      + \Or(x^3) \bigg\} \tilde{\psi}_A^{(1)} \nonumber\\
  &+ \Or(x^3) \tilde{\psi}_A^{(0)} \nonumber\\
  &= - \I \sigma^j D_j \tilde{\psi}_B^{(3)}
    + \left\{- \frac{mc^2}{6} \sigma^i R_{0lim} x^l x^m
      - \frac{m c^2}{12} \sigma^i R_{0lim;n} x^l x^m x^n
      + \Or(x^4)
    \right\} \tilde{\psi}_B^{(2)} \nonumber\\
  &- \sigma^j \bigg\{ \frac{\I}{2} a_j
      + \I (\vect a \cdot \vect x) D_j
      + \frac{\I c^2}{4} (R_{0j0l} - R_{jl}) x^l
      + \frac{\I c^2}{2} R_{0l0m} x^l x^m D_j
      + \frac{\I c^2}{6} \tensor{R}{^i_{ljm}} x^l x^m D_i \nonumber\\
  &\quad + \frac{\I c^2}{12} (R_{0j0l;m} - 2 R_{jl;m}) x^l x^m
      + \frac{\I c^2}{6} R_{0l0m;n} x^l x^m x^n D_j
      + \frac{\I c^2}{12} \tensor{R}{^i_{ljm;n}} x^l x^m x^n D_i \nonumber\\
  &\quad + \frac{\I c^2}{48} (R_{0j0l;mn} - 3 R_{jl;mn}) x^l x^m x^n
      + \frac{\I c^2}{24} R_{0k0l;mn} x^k x^l x^m x^n D_j
      + \frac{\I c^2}{40} \tensor{R}{^i_{kjl;mn}} x^k x^l x^m x^n D_i
    \nonumber\\
  &\quad + \Or(x^4) \bigg\} \tilde{\psi}_B^{(1)} .
\end{align}
Now we use \eqref{eq:PN_elim_psi_B_1}, \eqref{eq:PN_elim_psi_B_2},
\eqref{eq:PN_elim_psi_B_3} and \eqref{eq:PN_elim_psi_B_3_time_der} to
rewrite \eqref{eq:PN_psi_A_2} just in terms of $\tilde{\psi}_A$ and
read off the next-order Hamiltonian $H^{(2)}$ according to
\eqref{eq:Pauli_PN_2}:
\begin{align}
  H^{(2)}
  &= - \frac{(-\I \vect\sigma \cdot \vect D)}{4m^2}
        \{ \I q \vect\sigma \cdot \vect E
        + (-\I \vect\sigma \cdot \vect D) (H^{(0)} + q A_\tau) \}
    \nonumber\\
  &\quad - \frac{(-\I \vect\sigma \cdot \vect D)}{2m}
    \bigg\{m \vect a \cdot \vect x
      + \frac{mc^2}{2} R_{0l0m} x^l x^m
      + \frac{m c^2}{6} R_{0l0m;n} x^l x^m x^n
      - \I (\vect\omega \times \vect x)^i D_i \nonumber\\
  &\qquad + \frac{1}{2} \vect\sigma \cdot \vect\omega
      + \Or(x^4)
    \bigg\} \frac{(-\I \vect\sigma \cdot \vect D)}{2m} \nonumber\\
  &\quad - \frac{(-\I \vect\sigma \cdot \vect D)}{2m}
    \sigma^j \bigg\{ \frac{\I}{2} a_j
      + \I (\vect a \cdot \vect x) D_j
      + \frac{\I c^2}{4} (R_{0j0l} - R_{jl}) x^l
      + \frac{\I c^2}{2} R_{0l0m} x^l x^m D_j \nonumber\\
  &\qquad + \frac{\I c^2}{6} \tensor{R}{^i_{ljm}} x^l x^m D_i
      + \frac{\I c^2}{12} (R_{0j0l;m} - 2 R_{jl;m}) x^l x^m
      + \frac{\I c^2}{6} R_{0l0m;n} x^l x^m x^n D_j \nonumber\\
  &\qquad + \frac{\I c^2}{12} \tensor{R}{^i_{ljm;n}} x^l x^m x^n D_i
      + \frac{\I c^2}{48} (R_{0j0l;mn} - 3 R_{jl;mn}) x^l x^m x^n
      + \frac{\I c^2}{24} R_{0k0l;mn} x^k x^l x^m x^n D_j \nonumber\\
  &\qquad + \frac{\I c^2}{40} \tensor{R}{^i_{kjl;mn}} x^k x^l x^m x^n
        D_i
      + \Or(x^4) \bigg\} \nonumber\\
  &\quad- \sigma^j \bigg\{ \frac{\I}{2} a_j
      + \I (\vect a \cdot \vect x) D_j
      + \frac{\I c^2}{4} (R_{0j0l} - R_{jl}) x^l
      + \frac{\I c^2}{2} R_{0l0m} x^l x^m D_j
      + \frac{\I c^2}{6} \tensor{R}{^i_{ljm}} x^l x^m D_i \nonumber\\
  &\qquad + \frac{\I c^2}{12} (R_{0j0l;m} - 2 R_{jl;m}) x^l x^m
      + \frac{\I c^2}{6} R_{0l0m;n} x^l x^m x^n D_j
      + \frac{\I c^2}{12} \tensor{R}{^i_{ljm;n}} x^l x^m x^n D_i
    \nonumber\\
  &\qquad + \frac{\I c^2}{48} (R_{0j0l;mn} - 3 R_{jl;mn}) x^l x^m x^n
      + \frac{\I c^2}{24} R_{0k0l;mn} x^k x^l x^m x^n D_j
      + \frac{\I c^2}{40} \tensor{R}{^i_{kjl;mn}} x^k x^l x^m x^n D_i
    \nonumber\\
  &\qquad + \Or(x^4)
    \bigg\} \frac{(-\I \vect\sigma \cdot \vect D)}{2m}
\end{align}
Note that in the expression $- \frac{mc^2}{6} \sigma^i R_{0lim} x^l
x^m \tilde{\psi}_B^{(2)} = - \frac{c^2}{12} \sigma^i R_{0lim} x^l x^m
\{-\I \vect\sigma \cdot \vect D \tilde{\psi}_A^{(1)} + \Or(x^2)
\tilde{\psi}_A^{(0)}\}$ appearing in \eqref{eq:PN_psi_A_2}, the second
term is off our order of approximation, so we neglected it when
reading off $H^{(2)}$.  Explicitly evaluating the above expression, we
obtain the following order $c^{-2}$ Hamiltonian:
\begin{align} \label{eq:Pauli_H_PN_2}
  H^{(2)}
  &= -\frac{1}{4m} \vect a \cdot \vect D
    - \frac{\I}{4m} (\vect\sigma \times \vect a) \cdot \vect D
    - \frac{1}{2m} (\vect a \cdot \vect x)
      (\vect\sigma \cdot \vect D)^2
    - \frac{c^2}{4m} R_{0l0m} x^l x^m (\vect\sigma \cdot \vect D)^2
    \nonumber\\
  &\quad- \frac{c^2}{8m} R_{0l0m;n} x^l x^m x^n
      (\vect\sigma \cdot \vect D)^2
    - \frac{c^2}{24m} R_{0k0l;mn} x^k x^l x^m x^n
      (\vect\sigma \cdot \vect D)^2
    - \frac{1}{8m^3} (\vect\sigma \cdot \vect D)^4 \nonumber\\
  &\quad+ \frac{c^2}{8m} R + \frac{c^2}{4m} R_{00}
    + \frac{c^2}{12m} (4 \tensor{R}{^j_l} + \tensor{R}{_0^j_{0l}}) x^l
      D_j
    + \frac{\I c^2}{8m} \sigma^k
      (- 2 \tensor{\varepsilon}{^{ij}_k} R_{0l0i}
      + \tensor{\varepsilon}{^{im}_k} \tensor{R}{^j_{lim}})
      x^l D_j \nonumber\\
  &\quad- \frac{c^2}{6m} \tensor{R}{^i_l^j_m} x^l x^m D_i D_j
    + \frac{c^2}{16m} (R_{;l} + 2 \tensor{R}{^i_{l;i}}) x^l
    + \frac{\I c^2}{24m} \tensor{\varepsilon}{^{ij}_k} \sigma^k
      (R_{0i0l;j} - 2 R_{il;j}) x^l \nonumber\\
  &\quad+ \frac{c^2}{24m} \Big(5 \tensor{R}{^j_{l;m}}
        - 3 \tensor{R}{_0^j_{0l;m}}
        - \tensor{R}{_{0l0m}^{;j}}
        - \tensor{R}{^j_l^i_{m;i}}
        - \I \tensor{\varepsilon}{^{ij}_k} \sigma^k (2R_{0i0l;m}
          + R_{0l0m;i}) \nonumber\\
  &\qquad + 2\I \tensor{\varepsilon}{^{in}_k} \sigma^k
          \tensor{R}{^j_{lin;m}}\Big)
      x^l x^m D_j
    - \frac{c^2}{12m} \tensor{R}{^i_l^j_{m;n}} x^l x^m x^n D_i D_j
    \nonumber\\
  &\quad+ \frac{c^2}{48m} \Big(R_{;lm} + 4 \tensor{R}{^i_{l;im}}
        + \I \tensor{\varepsilon}{^{ij}_k} \sigma^k (R_{0i0l;jm}
          - 3 R_{il;jm})\Big)
      x^l x^m \nonumber\\
  &\quad+ \frac{c^2}{120m} \Big(9 \tensor{R}{^j_{l;mn}}
        - 6 \tensor{R}{_0^j_{0l;mn}}
        - 5 \tensor{R}{_{0l0m}^{;j}_n}
        - 3 \tensor{R}{^j_l^i_{m;in}}\Big)
      x^l x^m x^n D_j \nonumber\\
  &\quad+ \frac{\I c^2}{96m} \sigma^k \Big(
        {-4} \tensor{\varepsilon}{^{ij}_k} (R_{0i0l;mn} + R_{0l0m;ni})
        + 3 \tensor{\varepsilon}{^{ir}_k} \tensor{R}{^j_{lir;mn}}\Big)
      x^l x^m x^n D_j \nonumber\\
  &\quad- \frac{c^2}{40m} \tensor{R}{^i_k^j_{l;mn}} x^k x^l x^mx^n D_i
      D_j
    - \frac{q}{4m^2} \sigma^i \sigma^j D_i E_j
    - \frac{q}{12m} c^2 (R_{lm} + R_{0l0m}) x^l x^m
      \vect\sigma \cdot \vect B \nonumber\\
  &\quad+ \frac{q}{12m} \sigma^j c^2 R_{iljm} x^l x^m B^i
    + \frac{\I q}{4m^2} \vect\sigma \cdot (\vect\omega \times \vect B)
    + \frac{q}{2m^2} \vect\omega \cdot \vect B
    + \frac{q}{4m^2} (\omega_j x^i - \omega^i x_j)
      D_i B^j \nonumber\\
  &\quad+ \frac{\I q}{4m^2} (\vect\sigma \cdot
      (\vect\omega \times \vect x)) \vect B \cdot \vect D
    - \frac{\I q}{4m^2} \sigma^j
      (\vect\omega \times \vect x) \cdot \vect D B_j
    + \Or(x^3)
\end{align}
This is the final information that we need in order to calculate the
Pauli Hamiltonian up to and including the order of $c^{-2}$
\eqref{eq:Pauli_PN}.  Note that we have used the identity $\sigma^i
\sigma^j = \delta^{ij} \mathbb{1} + \I \tensor{\varepsilon}{^{ij}_k}
\sigma^k$ multiple times for simplifications, as well as the Bianchi
identities and $[D_i, D_j] = -\I q (\partial_i A_j - \partial_j A_i) =
-\I q \varepsilon_{ijk} B^k$, where $B^i = \varepsilon^{ijk}
\partial_j A_k$ is the magnetic field.  We also used that covariant
derivatives commute up to curvature terms, which are of higher order
in $c^{-1}$.

Note that due to a calculational oversight, the terms explicitly
containing the magnetic field were missing in the master's thesis
\cite{Alibabaei:2022Thesis} on which the present article is based.  In
\cite{Alibabaei:2022Thesis}, some oversights were also made regarding
the consistency of the calculation of the terms of order $x^2$.
However, the only differences of \eqref{eq:Pauli_H_PN_2} to the
corresponding result in \cite{Alibabaei:2022Thesis} that arise from
these miscalculations of order $x^2$ terms are the appearance of all
terms which contain covariant derivatives of the curvature tensor and
the absence of the term $- \frac{\I c^2}{4} (\vect\omega \times \vect
x)^k (R_{kl} + R_{0k0l}) x^l$ from \cite{Alibabaei:2022Thesis}.

\end{document}